\documentclass[11pt]{article}
\usepackage{amsmath}
\usepackage{float}
\usepackage{graphicx}
\title{}
\usepackage{listings}
\usepackage{color}
\usepackage{textcomp}
\definecolor{listinggray}{gray}{0.9}
\definecolor{lbcolor}{rgb}{0.9,0.9,0.9}
\usepackage[semicolon]{natbib}
\usepackage{mathpazo}
\usepackage{soul}
\usepackage{geometry}
\usepackage{mathtools}
\usepackage{textcomp}

\author{Niamh Cahill}
\emergencystretch=\maxdimen
\begin{document}

\title{A Bayesian Hierarchical Model for Reconstructing Sea Levels: From Raw Data to Rates of Change}
\author{Niamh~Cahill$^{1}$ Andrew C.~Kemp$^{2}$ Benjamin P~Horton$^{3,4}$ Andrew C.~Parnell$^{1}$\\ 
1. School of Mathematical Sciences, CASL, Earth Institute,
 University College Dublin\\ 
2. Dept. of Earth and Ocean Sciences, Tufts University\\ 
3. Department of Marine \& Coastal Sciences 
and Institute of Earth, Ocean, \\\& Atmospheric Sciences, Rutgers University\\
4. The Earth Observatory of Singapore, Nanyang Technological University}

\date{\today}

\maketitle

\begin{abstract}
We present a holistic Bayesian hierarchical model for reconstructing the continuous and dynamic evolution of relative sea-level (RSL) change with fully quantified uncertainty. The reconstruction is produced from biological (foraminifera) and geochemical ($\delta^{13}$C) sea-level indicators preserved in dated cores of salt-marsh sediment. Our model is comprised of three modules: (1) A Bayesian transfer function for the calibration of foraminifera into tidal elevation, which is flexible enough to formally accommodate additional proxies (in this case bulk-sediment $\delta^{13}$C values); (2) A chronology developed from an existing Bchron age-depth model, and (3) An existing errors-in-variables integrated Gaussian process (EIV-IGP) model for estimating rates of sea-level change. We illustrate our approach using a case study of Common Era sea-level variability from New Jersey, U.S.A. We develop a new Bayesian transfer function (B-TF), with and without the $\delta^{13}$C proxy and compare our results to those from a widely-used weighted-averaging transfer function (WA-TF). The formal incorporation of a second proxy into the B-TF model results in smaller vertical uncertainties and improved accuracy for reconstructed RSL. The vertical uncertainty from the multi-proxy B-TF is $\sim$28\% smaller on average compared to the WA-TF. When evaluated against historic tide-gauge measurements, the multi-proxy B-TF most accurately reconstructs the RSL changes observed in the instrumental record (MSE = 0.003 m$^2$). The holistic model provides a single, unifying framework for reconstructing and analysing sea level through time. This approach is suitable for reconstructing other paleoenvironmental variables using biological proxies.
 \end{abstract}

\newpage
\section{Introduction}
Paleoenvironmental reconstructions describe Earth's response to past climate changes and consequently offer a context for current trends and analogs for anticipated future changes (e.g., \citealp{Mann2009}). Reasoning by analogy underpins the use of biological proxies to reconstruct past environments (e.g., \citealp{Rymer1978, Jackson&WIlliams04, Bradley2015}). The ecological preferences of biological assemblages observed in modern environments are used to derive a paleoenvironmental reconstruction from their counterparts preserved in dated sediment cores under the assumption that the ecological preferences were unchanged through time  \citep{JugginsandBirks2012}. This approach commonly utilizes data consisting of one environmental variable and counts from multiple proxy species (e.g., \citealp{Imbrie1971, Fritz1991, Birks95}). Numerical techniques known as transfer functions formalize the relationship between biological assemblages and the environmental variable. This process is termed calibration. To quantify environmental change through time it is necessary to combine the paleoenvironmental reconstruction with a chronology of sediment deposition and an appropriate methodology to describe temporal trends. These three components can be developed and applied independently of one another or assimilated in a single, holistic framework.\\

Relative sea-level (RSL) reconstructions can constrain the relationship between temperature and sea level and reveal the long-term, equilibrium response of ice sheets to climate forcing (e.g., \citealp{Dutton2015}). Salt-marsh foraminifera are sea-level proxies, because species have different ecological preferences for the frequency and duration of tidal submergence, which is primarily a function of tidal elevation (e.g., \citealp{Scott78, Horton06, EdwardsandWright}). Under conditions of RSL rise, salt marshes accumulate sediment to maintain an elevation in the tidal frame. The resulting sedimentary sequence is an archive of past RSL changes that may be accessed by collecting sediment cores. After extraction, these sediment cores are sliced into layers (samples), from which foraminifera are counted. The transfer functions commonly used to reconstruct RSL impose a single ecological response to tidal elevation on all species of foraminifera (or other biological groups such as diatoms). Other analyses performed on the same layers can provide a multi-proxy approach to reconstructing RSL, although this often relies on informal approaches to combine results from independent proxies (e.g., \citealp{Kemp2013, Gehrels2000}). For example, on organogenic salt marshes on the U.S. Atlantic coast the primary source of organic carbon is \emph{in-situ} plant material and measurements of bulk sediment $\delta^{13}$C reflect the dominant plant community (e.g. \citealp{Kemp2012}). Some sediment layers are dated using radiocarbon or recognition of pollution markers of known age. Since there are typically fewer dated layers than total layers, a statistical age-depth model is used to estimate the age of undated layers with uncertainty (e.g., \citealp{OxCal, bchron, blaauw2011}).  Although Bayesian age-depth models and methods for estimating rates of sea-level change already exist, Bayesian methods are yet to be applied in the calibration phase of reconstructing RSL. This prevents the appropriate propagation of uncertainties, which is the primary advantage of using a holistic numerical framework.\\

We develop a Bayesian transfer function (B-TF) to reconstruct RSL using counts of foraminifera and measurements of bulk sediment $\delta^{13}$C from salt-marsh sediment.  This model allows each species of foraminifera to have a different ecological response to tidal elevation and provides a formalized approach to combine multiple proxies and consequently reduce reconstruction uncertainty. Following the framework of \citet{Bclim} we combine this new calibration module with an existing chronology module (Bchron), and an existing process module (the Errors-In Variables Integrated Gaussian Process (EIV-IGP) model of \citealp{Cahill15}) to create a holistic Bayesian hierarchical model. Through application of the model to a case study of Common Era and instrumental RSL change in New Jersey (USA), we compare the utility of the B-TF with an existing weighted averaging transfer function (WA-TF) approach and demonstrate the advantage of combining the three parts of a RSL reconstruction in a single and shared numerical framework rather than treating each as an independent and discrete step.

\section{Previous calibration methods}
Transfer functions are empirically-derived equations for reconstructing past environmental conditions from the abundance of multiple species. The term refers not to a single numerical method, but to a range of regression-based techniques that are classified into two categories depending on whether the underlying model maps environmental variables to species abundances (classical calibration) or vice versa (inverse calibration).  Classical approaches are underpinned by the ecologically-intuitive assumption that the distribution of species is driven by environmental variables \citep{Birks2012}. Inverse approaches gained popularity because of their reduced computational complexity (e.g., \citealp{PAGES}) resulting in quicker processing compared to classical methods.  Furthermore, inverse methods  often demonstrate equal or superior performance when compared to classical approaches (e.g., \citealp{Toivonen2000, ter1993, Korsman96}). The parameters in transfer functions are estimated using empirical data (a modern training set) from environments likely to be analogous to those encountered in core material (e.g., \citealp{JugginsandBirks2012}) and are treated as fixed and known.  Studies seeking to reconstruct RSL from salt-marsh sediment employ transfer functions developed using a modern training set of paired observations of tidal elevation and microfossil assemblages (most commonly foraminifera or diatoms) to reconstruct RSL from their counterparts preserved in sediment cores (e.g., \citealp{Horton1999, Gehrels2000, EdwardsandHorton2006, Kemp13, Barlow2014}).  Although the different types of transfer function have advantages and weaknesses compared to one another, these regression-based techniques share the limitations of applying a single response form to all species and treating model parameters as fixed and known.  These characteristics can result in misleading or inaccurate paleoenvironmental reconstructions if the response curve is not appropriate for all species \citep{Smith83} and does not account for the inherent uncertainty in model parameters that results from ecological noise and the influence of secondary environmental variables, which in RSL reconstructions can include salinity and sediment texture and composition (e.g., \citealp{Shennan1996, Zong1999}). \\

Bayesian calibration methods are inherently classical and have recently been given growing attention to produce paleoenvironmental reconstructions using biological proxies (e.g., \citealp{Toivonen2000, Bummer, Haslett2006, Bo2010, Tingley2012, Tolwinski2013, Tolwinski2015, Bclim}). \citet{Toivonen2000} and \citet{Bummer} developed a Bayesian model to reconstruct temperature from chironomid counts. \citet{Haslett2006} adopted elements of the model proposed by \citet{Toivonen2000}  in a more complex Bayesian hierarchical model for reconstructing multivariate climate histories from pollen counts. \citet{Bo2010} proposed a Bayesian hierarchical model to reconstruct temperature using a multi-proxy approach. Similarly, \citet{Tingley2012} considered a Bayesian hierarchical space-time model for inferring climate processes. More recently, \citet{Tolwinski2013, Tolwinski2015} and \citet{Bclim}  expanded on the aforementioned approaches of \citet{Haslett2006}  and \citet{Tingley2012} for reconstructing climate variables. To date, Bayesian methods have not been used for reconstructing RSL using biological proxies.

\section{A Bayesian hierarchical model for reconstructing and analysing former sea levels}
\label{method1}
We now describe our statistical model, which produces estimates of RSL and associated rates from raw inputs including foraminifera counts and radiocarbon dates from a sediment core. We add two major novelties to existing approaches:
\begin{enumerate}
\item  A B-TF model using a penalized spline (P-spline) as a non-parametric model of the multinomial response of foraminifera to tidal elevation. This model allows for multi-modal and non-Gaussian species response to environmental variables;
\item A full hierarchical model which incorporates the B-TF, a chronology model accounting for time uncertainty, and a rich stochastic process for quantifying sea level rate changes.
\end{enumerate}
We use the JAGS package (Just Another Gibbs Sampler; \citealp{JAGS}) to fit the model via Gibbs sampling.\\

We start by outlining our notation:
\begin{itemize}
\item $y$ are the observed foraminifera abundances from the sediment core. $y_{il}$ is the abundance of species $l$ in layer $i$. We denote $y_i$ the L-vector of foraminifera counts for each layer $i$ in the sediment core, where $i=1,\ldots,N$ layers and $l=1,\ldots,L$ species;
\item $r$ are the observed radiocarbon dates in the sediment core. $r_k$ is the $k^{th}$ radiocarbon date, $k=1,\ldots,K$. Usually $K \ll N $. Due to the nature of radiocarbon, these are given in radiocarbon years rather than calendar years. A known calibration curve is used to transform the radiocarbon ages into calendar ages as part of the chronology model (Sect. \ref{chron});
\item $d$ are the observed depths in the sediment core. $d_i$ is the depth associated with layer $i$;
\item $e$ is paleo marsh elevation (PME), which is the tidal elevation at which a layer originally accumulated. $e_i$ is the PME for sediment core layer $i$;
\item $s$ is RSL. $s$ has a deterministic relationship with $e$ and $d$ given some fixed parameters $\omega$ so that $s = g_{\omega}(e, d)$. Producing $s$ will require correcting PME for sample tidal elevation (a function of sediment core depth). $\omega$ includes values for the the sample tidal elevation (E) so that $s_i=E_i - PME_i$. $s_i$ is the RSL for sediment core layer $i$;
\item $t$ represents the calendar ages (in years before present (1950); BP) of all layers in the sediment core. It is unknown and estimated with uncertainty as part of the chronology module from the radiocarbon dates $r$ and observed depths $d$. $t_i$ represents the age of sediment core layer $i$;
\item $y^{m}$ are the observed modern foraminifera counts. $y^m_{jl}$ is the abundance of species $l$ in surface sample $j$. $y^m_j$ is an L-vector of modern foraminifera counts for modern sample $j$ with $j=1,\ldots, J$ modern samples. $T_j = \sum_{l=1}^L y^m_{jl}$ are the row totals of species counts for calibration sample $j$ in the matrix of species abundances;
\item $e^m$ are the observed modern tidal elevations. $e^m_j$ is the tidal elevation for surface sample $j$. Together $y^m$ and $e^m$ are used to calibrate the relationship between  foraminifera abundance and tidal elevation;
\item $z$ is the sediment core $\delta^{13}$C where $z_i$ is the $\delta^{13}$C for layer $i$. We include this as a secondary proxy though it is an optional part of the model and can be removed if unavailable in other sediment cores;
\item $\theta$ are a set of parameters governing the relationship between foraminifera counts and tidal elevation;
\item $\psi$ are a set of parameters governing the sedimentation process (i.e. linking age and depth);
\item $\phi$ are a set of parameters governing the RSL process, including its smoothness and variability;
\item $\alpha$ are a set of parameters governing the relationship between $\delta^{13}$C and tidal elevation.
\end{itemize}

Using the notation above we create a Bayesian hierarchical model to produce a posterior distribution of our parameters given data:
\begin{center}
$p(s,e,t,\theta, \psi, \phi,\alpha|y^m,e^m,y,r,d,\omega,z) \propto$\\
\end{center}
\begin{center}
$ \underbrace{p(y|e,\theta)p(z|e,\alpha)}_\text{Fossil Data Model} \times \underbrace{p(y^m|e^m, \theta)}_\text{Modern Calibration Model} \times \underbrace{p(r|t,\psi,d)}_\text{$^{14}$C Calibration} \times \underbrace{p(t|d,\psi)}_\text{Chronology Model} \times \underbrace{p(s|e,d,t,\phi,\omega)}_\text{Sea level Model}$\\
\end{center}
\begin{center}
$\times \underbrace{p(\theta)}_\text{Modern Calibration Prior} \times \underbrace{p(\psi)}_\text{Chronology Prior} \times \underbrace{p(\phi)}_\text{Sea Level Prior} \times \underbrace{p(\alpha)}_\text{$\delta^{13}$C Prior} $
\end{center}
Before describing the components of the model that we use, we note that this is an extremely complex and computationally demanding model to fit, being of very high dimension with rich stochastic processes being required for many of the sub-models. We follow \citet{Bclim} in making some simplifying assumptions. We first assume that the calibration parameters $\theta$ can be learnt solely from the modern calibration data $y^m$ and $e^m$. Thus the sediment core data contains no further information about this relationship. This is a common assumption in many palaeoclimate studies (see e.g. \citealp{Haslett2006,Ward2015}). Second we assume that the model can be modularised into three parts: the aforementioned calibration, chronology and process modules. This is a conservative assumption and follows from the restriction on the calibration parameters.\\

Following these assumptions we obtain the three modules:
\begin{align*}
p(t,\psi|r,d) &\propto p(r|d,t,\psi) p(t|d,\psi) p(\psi) \tag{chronology module}\\[0.2cm]
\vspace{0.2cm}
p(\theta|y^m,e^m) & \propto p(y^m|e^m,\theta) p(\theta) \tag{calibration module}\\[0.2cm]
p(s,e,t,\theta,\psi,\phi, \alpha|y^m,e^m,y,r,d,\omega,z) &\propto p(y|e,\theta) p(\theta|y^m,e^m) p(t,\psi|r,d) p(s|e,\omega,d,t,\phi) p(\phi) p(z|e,\alpha)p(\alpha) \tag{process module}
\end{align*}
We note that if there is no additional $\delta^{13}C$ proxy information then $z$ and $\alpha$ (and hence the last two terms on the RHS of the process module) are removed from the equation.

\subsection{The calibration module: multinomial P-splines (B-TF)}
\label{multimodel}
In this module we aim to estimate the parameters $\theta$ that govern the relationship between forami-nifera and tidal elevation by using the model as specified in the previous section. The probability density function (pdf) $p(y_j^m|e_j^m,\theta)$ used as the likelihood here provides the data-generating mechanism from which foraminifera abundances can be simulated given tidal elevation. The likelihood we use for the modern model is:

\begin{align}
y^m_{j1},y^m_{j2},...y^m_{jl}|T_j,p_{j1},p_{j2},....p_{j,l} \sim Multinomial(p_{j1},p_{j2},....p_{jl}, T_j),
\end{align}

\noindent where $p_j=\left\{p_{j1},p_{j2},....p_{jl}\right\}$ is the vector containing the probability of finding species $l$ at the tidal elevation associated with sample $j$.\\

The probability vectors $p_j$ are estimated from a latent response $\lambda_{jl}$ (i.e. the response of species $l$ for sample $j$) which is a function of tidal elevation $e^m_j$. $\lambda_{l}$ is a J-vector including the latent response of species $l$ for all samples $j$. The relationship between probability of foraminifera species occurrence and tidal elevation is expected to be non-linear so we model these using P-splines \citep{deboor78,Dierckx93} via a softmax transformation. The softmax transformation is given as:

\begin{align}
p_{jl} &=\frac{\exp(\lambda_{jl})}{\sum_{l=1}^L \exp(\lambda_{jl})}
\end{align}

The $\lambda$ parameters are given $P$-spline prior distributions. P-splines are created from B-spline basis functions penalised to produce a smooth curve. The B-spline basis functions are constructed from piecewise polynomial functions that are differentiable to a given degree $q$, here cubic. The component cubic B-spline basis functions look like individual Gaussian curves, however, they will be non-zero only over the range of $q+2$ knots; this has numerous computational advantages. We refer to the B-spline matrix as $B$. The columns of $B$ are the tidal elevations $e^m$, transformed by the appropriate basis function. The resulting relationship is:

\begin{align}
\lambda_{l} &=B\beta_l+\epsilon_l
\end{align}

$B$ is a $J\times M$ matrix of basis functions where $M$ is the number of knots, and $J$ is the number of modern samples. To obtain the penalised smooth behaviour for $\lambda$ we apply a prior such that the first differences of $\beta_l$ are normally distributed with mean 0 and precision $\tau_{\beta}$. The parameter $\tau_{\beta}$ controls how close the weights are related to each other and will therefore control smoothness. \\

An error term, $\epsilon_l \sim N(0,\nu_l)$, is added to the mean for $\lambda$ here to ensure that we do not encounter problems with over-dispersion by under or over-estimating the variance in the observed data.We do not assume a constant variance; to account for the changing variation in the data the precision parameters $\nu_l$ are also estimated using P-splines. This allows the variance to adapt given the data and will allow it to increase/decrease where necessary.

\begin{align}
\nu_l &=exp(B\gamma_l)
\end{align}

Similarly to $\lambda^m$, the basis functions are penalised by parameters $\gamma_l$ to produce $\nu$ and we apply a prior such that the first differences of $\gamma_l$ are normally distributed with mean 0 and precision $\tau_{\gamma}$. Therefore, the calibration model has parameters $\theta = \left\{ \beta_l, \gamma_l, \lambda , \tau_{\gamma}, \tau_{\beta}; \hspace{0.5em} l=1,\ldots ,L \right\}$,  which can be fitted in a single Bayesian model for all species simultaneously. \\

The B-TF produces posterior estimates for the multinomial probability vector $p$ for each modern sample. For each species of foraminifera, we compare the probability of species occurrence (at each modern observed tidal elevation) estimated from the B-TF with the empirical probability of foraminifera species occurrence estimated from the observed data. The model vs. empirical probability comparison provides evidence to support the validity of the model, indicating if the model is capable of capturing the within-species variability of occurrence probabilities across changing tidal-elevations. Once run, the B-TF can produce predictions of elevation for each layer in the sediment core from this relationship.\\

We evaluate the performance of the B-TF via 10-fold cross validation on the modern data, where the data are divided up into 10 randomly drawn equal size sections (known as folds) which are removed in turn. We create predictions for the left out sections repeatedly until every observation has an out-of-sample prediction value.To allow direct and meaningful comparison between models we also cross validated the WA-TF using the same approach on the same randomly drawn folds. We showcase the output of this exercise for our case study in Sect \ref{CV}.

\subsection{The chronology module: Bchron}
\label{chron}
The chronology module is concerned with estimating the ages $t$ of the foraminifera in the sediment core. These ages will necessarily be uncertain, since the radiocarbon dates $r$ are observed with uncertainty which, when transformed into calendar years, provide highly non-Gaussian probability distributions. An interpolation step is then required to obtain estimated ages at all depths, which adds further uncertainty. A useful constraint is that age must increase with depth (older sediments lie deeper in the core, known as superposition) so a monotonic stochastic process is used. Bchron \citep{bchron} assumes that the integrated sedimentation rate (i.e. the accumulation of sediment over a fixed period of time) arises as the realisation of a Compound Poisson-Gamma (CPG) process. Bchron calibrates the radiocarbon (and non-radiocarbon) dates, estimates the parameters of the CPG (here $\psi$) and identifies outliers. Other age-depth models are available (see \citealt{Parnell2011} for a review), but Bchron was designed specifically for use in palaeoenvironmental reconstructions. \\

Once Bchron has been run, we obtain a joint posterior distribution of ages for every layer in the sediment core, which we denote as $p(t|r,d,\psi)$. Each individual chronology sample from Bchron satisfies the law of superposition. However, we approximate the age of each layer in the posterior, i.e. $p(t_i|r,d,\psi)$ as a normal distribution, so that $t_i|r,d,\psi \hat{\sim} N(\mu_{t_i},\sigma^2_{t_i})$. This may seem like a severe relaxation, since the ages of layers may now overlap, but we find this has minimal effect on the resulting sea level curves since the ages are further updated during the process module. Further simulations justifying this assumption have been carried out using chronological models in late Holocene sea level reconstructions from saltmarsh sediments \citep{ParnellandGehrels}. 

\subsection{The process module: errors-in-variables integrated Gaussian process (EIV-IGP)}
Our final step is to take the output from the previous two modules, namely estimates of the posterior PME $e_i$ for each sediment core layer from the calibration module, and estimates of the age of each layer $t_i$ from the chronology module. In cases where the secondary $\delta^{13}$C proxy is available, the posterior estimated for $e_i$ will include the likelihood $p(z|e,\alpha)$. This is a normal likelihood $z_i \sim N(\mu_i,\tau_z)$ where the precision $\tau_z$ is constant and $\mu_i$ will correspond to the dominant $\delta^{13}$C value at $e_i$. $\delta^{13}$C reflect dominant plant communities on a marsh and the observed modern boundries between communities can correspond to a tidal datum (TD). As a result $\delta^{13}$C measured in bulk sediment can be related to tidal elevation as follows;

\[
    \mu_i= 
\begin{cases}
    \mu_1,& \text{if } e_i\leq TD\\
    \mu_2,& \text{if } e_i\geq TD\\
    \mu_3,              & \text{otherwise}
\end{cases}
\]

$\mu_i$'s are given informative uniform priors with upper and lower limits corresponding to the maximum and minimum $\delta^{13}$C values represented in a given elevational range. The prior information required here is location specific. The details needed for priors related to our case study are presented in Section \ref{proxies}. \\

We can transform $e_i$ into RSL $s_i$ via the relationship $s_i = g_\omega(e_i,d_i)$. We thus have a set of bivariate probability distributions for each layer consisting of pairs $(t_i,s_i)$ which represent the raw layer-by-layer estimates of RSL and age. To use these in the EIV-IGP framework of \citet{Cahill15}, we approximate each bivariate probability distribution as bivariate Gaussian. The model makes use of two well known statistical approaches. Firstly, the EIV approach \citep{EIV} accounts for measurement error in the explanatory variable, here time. The EIV approach is necessary when dealing with proxy reconstructions that include temporal uncertainty from dating the sediment core. Secondly, the Gaussian process approach \citep{Rasmussen06GP} is useful for nonlinear regression problems and is a practical approach to modelling time series data. A Gaussian process is fully specified by a mean function (set to zero) and a covariance function that relates the observations to one another. \\

We use an integrated Gaussian process approach \citep{Holsclaw2013, Cahill15}. A Gaussian process prior is placed on the rates of sea-level change and the mean of the distribution assumed for the observed data is derived from the integral of the rate process. This integrated approach is useful when there is interest in the rate process as the analysis allows for estimates of instantaneous rates of sea-level change. Furthermore the current sea level estimate is derived as the integral of all the previous sea level rates that have occurred, matching the physical behaviour of sea level evolution over time. By embedding the integrated Gaussian process (IGP) model in an errors-in-variables (EIV) framework (which takes account of time uncertainty), we can estimate rates with quantified uncertainty. We use the same priors for the parameters $\phi$ as described in Cahill et al. (2015) where technical details of the IGP-EIV model can be found.

\section{Case study: New Jersey RSL}
On the Atlantic coast of southern New Jersey (Figure 1), salt marshes form in quiet-water, depositional settings and display a zonation of plants into distinct vertical zones corresponding to ecologically important tide levels. Elevations below mean tide level (MTL) are not vegetated and the inorganic sediment is comprised of silt and fine sand with shell material. Low salt-marsh environments between MTL and mean high water (MHW) are vegetated by \emph{Spartina alterniflora} (tall form), which is a C$_4$ plant. Sediment in this zone is organic grey silt and clay. High salt-marsh environments exist between MHW and highest astronomical tide (HAT). This zone is typically a wide, flat meadow vegetated by \emph{Spartina patens} and \emph{Distichlis spicata} (C$_4$ species). The sediment deposited in this zone is brown peat with abundant plant remains. The transition between high salt marsh and the freshwater upland is vegetated by C$_3$ plants such as \emph{Phragmites australis}, \emph{Iva fructescens}, \emph{Schoeneplectus americanus}, and \emph{Typha augusitfolia}. This community exists at tidal elevations above mean higher high water (MHHW), including freshwater environments above the reach of tidal influence, and occurs with black, amorphous, organic sediment.

\begin{figure*}[h!]
\begin{center}
\includegraphics[scale=1.1]{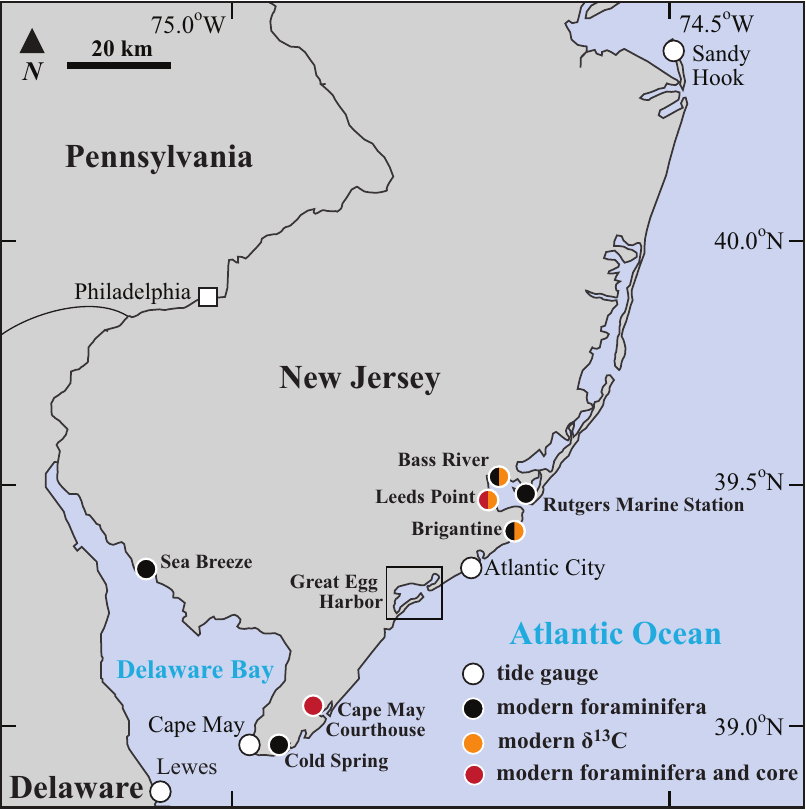}
\caption{Location of study sites in southern New Jersey, USA. The distribution of modern foraminifera was described at 12 different salt marshes including five in Great Egg Harbor (not located with symbols in the figure).  Bulk surface sediment $\delta^{13}$C values were measured at three sites and cores for sea-level reconstruction (filled circles) were collected at Leeds Point and at Cape May Courthouse.}
\end{center}
\end{figure*}

\subsection{Modern training set}
At twelve salt marshes in southern New Jersey \citet{Kemp2013} established transects across the prevailing environmental gradient from lower to higher tidal elevations (Figure 1). The twelve sites were selected to span a wide range of physiographic settings including brackish marshes located up to 25 km from the coast with a strong fluvial influence. The sites share a common climate and oceanographic regime and therefore constitute a regional-scale training set. At stations along each transect a surface sediment sample was collected to describe the assemblage of foraminifera (count sizes ranged from 8 to 307 dead individuals). The tidal elevation of each sample was measured in the field. \\

Since the great diurnal tidal range (MLLW to MHHW) varies among sites in the study region it is necessary to express tidal elevation as a standardized water level index (SWLI; e.g. \citealp{Horton1999}), where a value of 0 corresponds to MLLW and 100 is MHHW. At NOAA tide gauges in New Jersey, measured HAT occurs at SWLI values of 127 in Atlantic City and 123 at Cape May.

\subsubsection{Modern counts of foraminifera}
The modern dataset comprised of 172 paired observations of 18 foraminiferal species (including many zeros) and tidal elevation. The highest occurrence of foraminifera in the modern dataset is 141.5 SWLI. Higher samples were devoid of foraminifera and interpreted as being from a freshwater environment above marine influence. This modern training set demonstrates that foraminifera (like plants) form distinct assemblages that correspond to elevation in the tidal frame (e.g., \citealp{Scott78}), but with a secondary influence of salinity (e.g., \citealp{deRijk1995}). Throughout southern New Jersey, low-marsh environments are occupied by \emph{Miliammina fusca} and \emph{Ammobaculties spp.} (groups D and E in Figure 2). High salt-marsh environments are characterized by a number of foraminiferal assemblages including groups dominated by \emph{Trochammina inflata}, \emph{Arenoparella mexicana}, and \emph{Tiphotrocha comprimata}. High salt marshes at sites with strong fluvial influence and correspondingly low (brackish) salinity are occupied by \emph{Ammoastuta inepta} (group G in Figure 2). At some sites elevations above MHHW are characterized by a group of foraminifera in which \emph{Haplophragmoides manilaensis} is the dominant species (group A in Figure 2). The pattern (uniform low marsh and diverse high marsh) and composition of these assemblages is similar to those identified elsewhere on the U.S. Atlantic coast (e.g., \citealp{Murray1991, Gehrels1994, Kemp09, Wright2011, Edwards2004}). This modern training set, was previously used to develop a WA-TF \citep{Kemp2013}, and is also used to develop our B-TF.

\subsubsection{Modern bulk-sediment $\delta^{13}$C measurements}
In the mid-Atlantic and northeastern U.S. the low salt-marsh and high salt-marsh zones are dominated by C$_4$ species such as \emph{Spartina alterniflora}, \emph{Spartina patens}, and \emph{Distichlis spicata}, while the transitional marsh and surrounding upland zones are dominated by C$_3$ species. In New Jersey the boundary between C$_3$ and C$_4$ plant communities corresponds to MHHW and $\delta^{13}$C measured in bulk sediment can be used to reconstruct RSL by determining if a sample formed above or below the MHHW tidal datum.\\

Based on the modern dataset of bulk sediment $\delta^{13}$C from three sites in southern New Jersey (Figure 1) and presence or absence of foraminifera,  \citet{Kemp2013} recognized three types of sediment that were likely to be encountered in cores of organic coastal sediment. 

\begin{enumerate}
\item Samples with $\delta^{13}$C values more depleted than -22.0\textperthousand\  and with foraminifera present formed at tidal elevations from 100-150 SWLI. The lower limit of this range corresponds to MHHW and the upper limit is conservatively set to extend slightly beyond the observed highest occurrence of foraminfera (141.5 SWLI) in the modern dataset.

\item Samples with $\delta^{13}$C values less depleted than -18.9\textperthousand\ formed at tidal elevations from 0-100 SWLI since C$_4$ plants are dominant below MHHW. This interpretation is the same if foraminifera are present or absent.

\item Samples with intermediate $\delta^{13}$C values between -18.9\textperthousand\ and -22.0\textperthousand\ provide no additional information and if foraminifera are present these samples are interpreted as having formed at 0-150 SWLI (MLLW to slightly above the highest observed occurrence of foraminifera).
\end{enumerate}

\begin{figure}[h!]
\begin{center}
\includegraphics[scale=0.88]{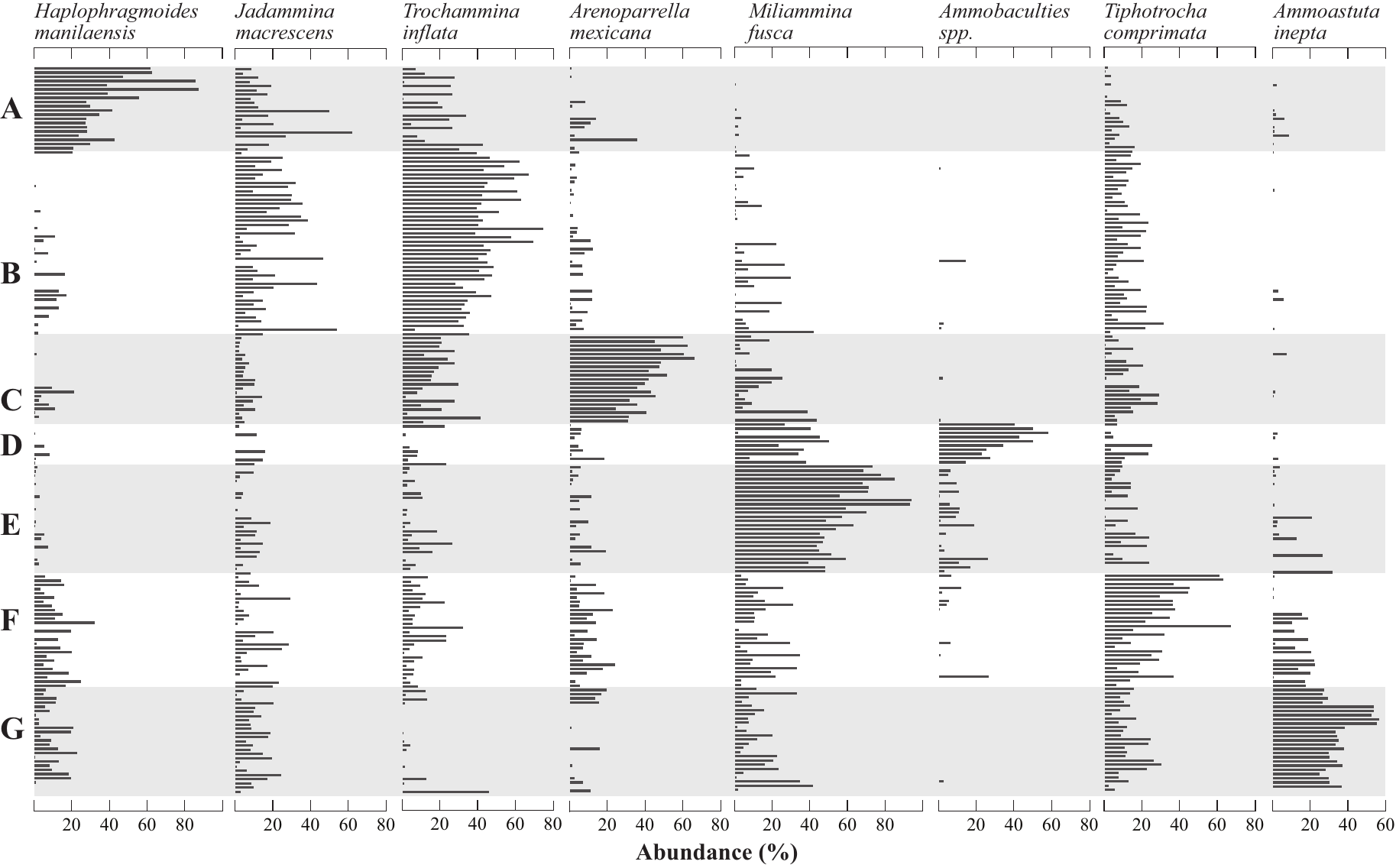}
\caption{Dataset of modern foraminifera described from a total of 172 surface sediment samples from 12 different sites. The samples were grouped using partioning around medoids. Only the abundance of the eight most common species are shown. Modified from Kemp et al. (2013).}
\end{center}
\end{figure}

\subsection{Proxy data}
\label{proxies}
Cores of salt-marsh sediment were recovered from two sites in southern New Jersey (Cape May Courthouse and Leeds Point; Figure 1) and sliced into 1-cm thick samples. Three types of data were generated for each sediment core and were originally presented by \citet{Kemp13}.

\subsubsection{Fossil counts of foraminifera}
In the Cape May Courthouse core \emph{Jadammina macrescens} and \emph{Trochammina inflata} were the dominant species from 1.72 m to 1.29 m (Figure \ref{ResultsFig}A, upper panel). Foraminifera were absent at 1.25 m to 1.12 m. Between 1.10 m and 0.33 m \emph{Jadammina macrescens} was the dominant species, while samples in the interval from 0.31 m to 0.05 m included \emph{Trochammina inflata}, \emph{Tiphotrocha comprimata}, and \emph{Jadammina macrescens}. Two samples near the top of the core (0.03 m and 0.05 m) included 17\% and 21\% \emph{Miliammina fusca} respectively. Counts of foraminifera in this core ranged from 16 to 194 per sample with an average of 98. In the lower part of the Leeds Point core (3.95 m to 2.85 m) \emph{Jadammina macrescens} was the most common species and occurred with \emph{Tiphotrocha comprimata} and \emph{Trochammina inflata} (Figure \ref{ResultsFig}A, lower panel). Within this section there were unusual occurrences of \emph{Miliammina petila} (24-60\% at 3.13 m to 3.30 m) and \emph{Miliammina fusca} ($>$20\% from 2.82 m to 2.95 m). From 2.82 m to 1.85 m \emph{Trochammina inflata} was the dominant species. The uppermost section of the Leeds Point core (1.73 m to 1.20 m) was comprised of a near mono-specific assemblage of \emph{Jadammina macrescens}. Counts of foraminifera in this core ranged from 4 to 127 per sample with an average of 70. For both cores the preserved assemblages of foraminifera were compared to the composition of modern samples in the training set. If core samples exceeded the 20$^{th}$ percentile of dissimilarity measured using the Bray-Curtis metric among all possible pairings of modern samples then the core sample was deemed to lack a suitable modern analogue and was excluded from further analysis by \citet{Kemp13}.  We did not reconstruct PME for these samples because they may lack ecological plausibility (e.g., \citealp{Jackson&WIlliams04}). On Figure \ref{ResultsFig} these samples are lacking PME reconstructions (panels B, C, and E).

\subsubsection{Fossil bulk-sediment $\delta^{13}$C measurements}
In the Cape May Courthouse core all samples were less depleted than -18.9\textperthousand\ and were interpreted as having formed below MHHW in a salt marsh dominated by C$_4$ plants (Figure \ref{ResultsFig}D, upper panel).  In the lowermost part of the Leeds Point core (below 3.35 m) the presence of foraminifera and bulk-sediment $\delta^{13}$C values more depleted than -22.0\textperthousand\ indicate that the sediment accumulated above MHHW in an environment dominated by C$_3$ plants, but below the highest occurrence of foraminifera (Figure \ref{ResultsFig}D, lower panel).  Between 3.31 m and 2.86 m bulk-sediment $\delta^{13}$C values were variable and interpreted to record the transition from highest salt marsh to high salt-marsh environments.  Above 2.86 m, all samples were less depleted than -18.9\textperthousand\ and were interpreted as having formed below MHHW in a salt marsh dominated by C$_4$ plants.

\subsubsection{Age-depth profile estimated by Bchron}
\begin{figure}[h!]
\begin{center}
\includegraphics[scale=0.75]{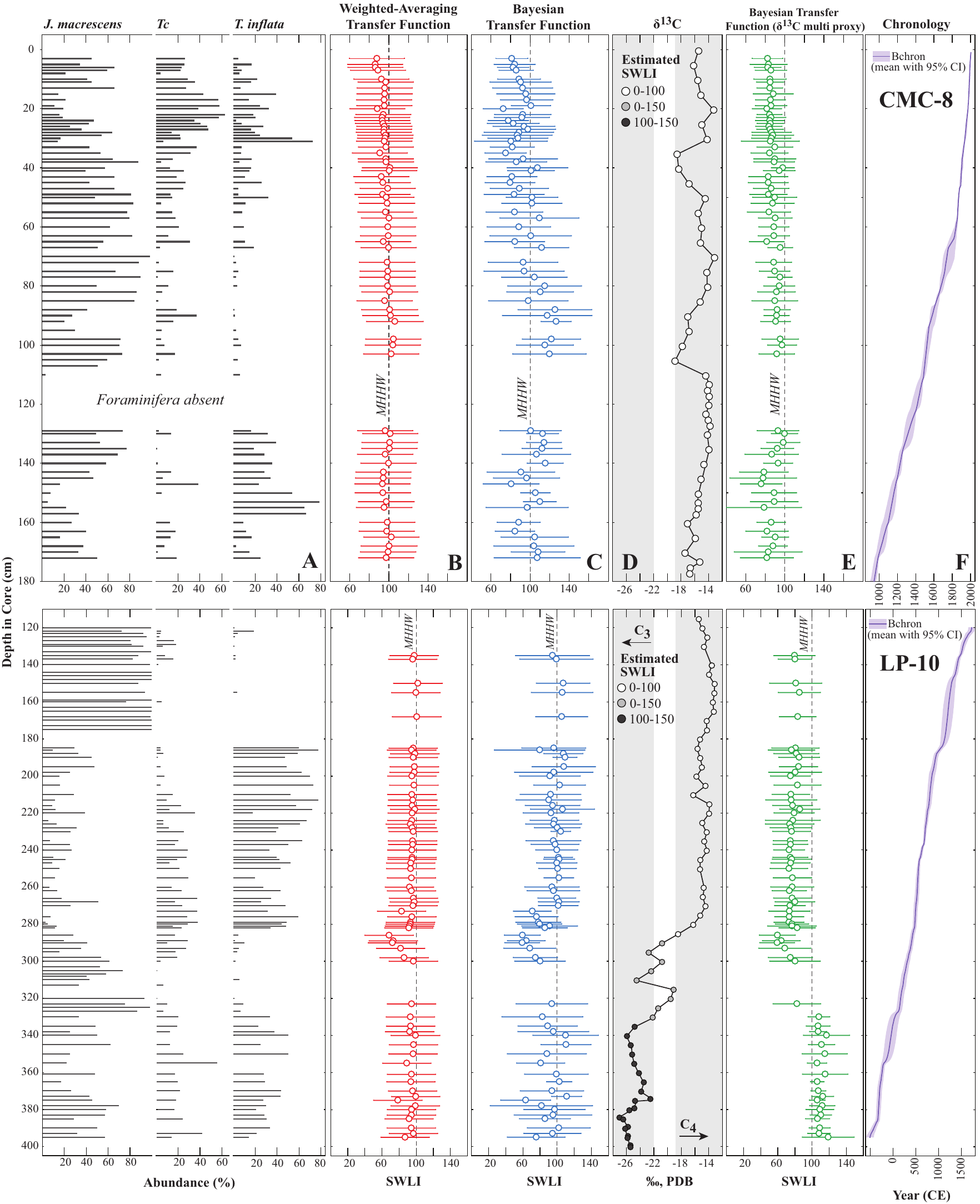}
\caption{The abundance of the the three most common species (Tc= \emph{Tiphotrocha comprimata}) of foraminifera found in cores from Cape May Courthouse (CMC-8) and Leeds Point (LP-10) are represeted by the horizontal gray bars (A). Paleo marsh elevation, in standardised water level index (SWLI) units, was reconstructed using the weighted average transfer function (B), the Bayesian transfer function (C) and the multi-proxy Bayesian transfer function (E). Mid points of the reconstruction are shown as white circles with the bars representing $\pm$2$\sigma$ uncertainty. Vertical dashed lines show the elevation of the mean higher high water (MHHW) tidal datum. Stable carbon isotope concentrations ($\delta^{13}$C) for bulk sediment are parts per thousand (\textperthousand) relative to the Vienna Pee Dee Belemnite (VPDB) standard (D). Bchron provided the chronology for both cores (F). Note that some samples with counts of foraminifera lack a corresponding PME reconstruction because they lack a suitable modern analog using the criteria applied by Kemp et al. (2013).}
\label{ResultsFig}
\end{center}
\end{figure}

Age-depth models for the Cape May Courthouse and Leeds Point cores were previously developed using Bchron \citep{Kemp2013} and are used here in the chronology module without modification (Figure \ref{ResultsFig}F). The Cape May Courthouse core was dated by recognition of pollution markers, the appearance of Ambrosia pollen as a land clearance marker, and radiocarbon dating of \emph{in situ} and identifiable plant macrofossils. These data were combined into a single age-depth model that estimated the age of every 1-cm thick sediment sample in the core with an average uncertainty of $\sim$30 years for the period since $\sim$700 CE (Figure \ref{ResultsFig}F, upper panel).   Anthropogenic modification of the Leeds Point site limited dating and RSL reconstruction to the interval from $\sim$500BC to $\sim$1750 CE. The core was dated using only radiocarbon measurements performed on \emph{in situ} and identifiable plant macrofossils. These data were combined into a single age-depth model that estimated the age of every 1-cm thick sediment sample in the core with an average uncertainty of $\sim$50 years (Figure \ref{ResultsFig}F, lower panel). 

\subsection{Instrumental data}
A tide gauge is an instrument that automatically measures the sea surface height with reference to a control point on the land many times during a day. These measurements are averaged to obtain annual values to minimize the effects of weather and tidal variability. In New Jersey tide-gauge data are available since 1911 CE when the Atlantic City tide gauge was installed. The Sandy Hook, Cape May, and Lewes (Delaware) tide gauges began measurements in 1932 CE, 1966 CE, and 1919 CE respectively. A single regional record was compiled by averaging annual data from these four local tide gauges. RSL is zero between 2000-2010 CE to be roughly equivalent to the age of a surface sample in the core (by definition RSL=0 m when the core was collected). The resulting record minimizes spatial variability and the influence of decadal-scale RSL variability \citep{Douglas91}. A linear regression of the averaged record shows that RSL rose at an average rate of 4.03 mm/yr between 1911 CE and 2012 CE \citep{Kemp2013}.

\section{Results}
We reconstruct PME using the original WA-TF of \citet{Kemp2012,Kemp2013} and our new B-TF. We developed a third reconstruction by incorporating downcore $\delta^{13}$C values with our B-TF to inform the posterior distribution for PME (see Section 3: process module). These results are combined with the existing Bchron age-depth model for each sediment core to reconstruct RSL. The resulting reconstructions are analysed using the EIV-IGP model to capture the continuous and dynamic evolution of RSL change while taking account of uncertainty in both sea level and age reconstructions. Our goal is honest assessment of uncertainty rather than reduced uncertainty.

\subsection{The Bayesian Transfer Function}
\subsubsection{Species-response curves}
\label{RC}

\begin{figure}[h!]
\begin{center}
\includegraphics[scale=0.36]{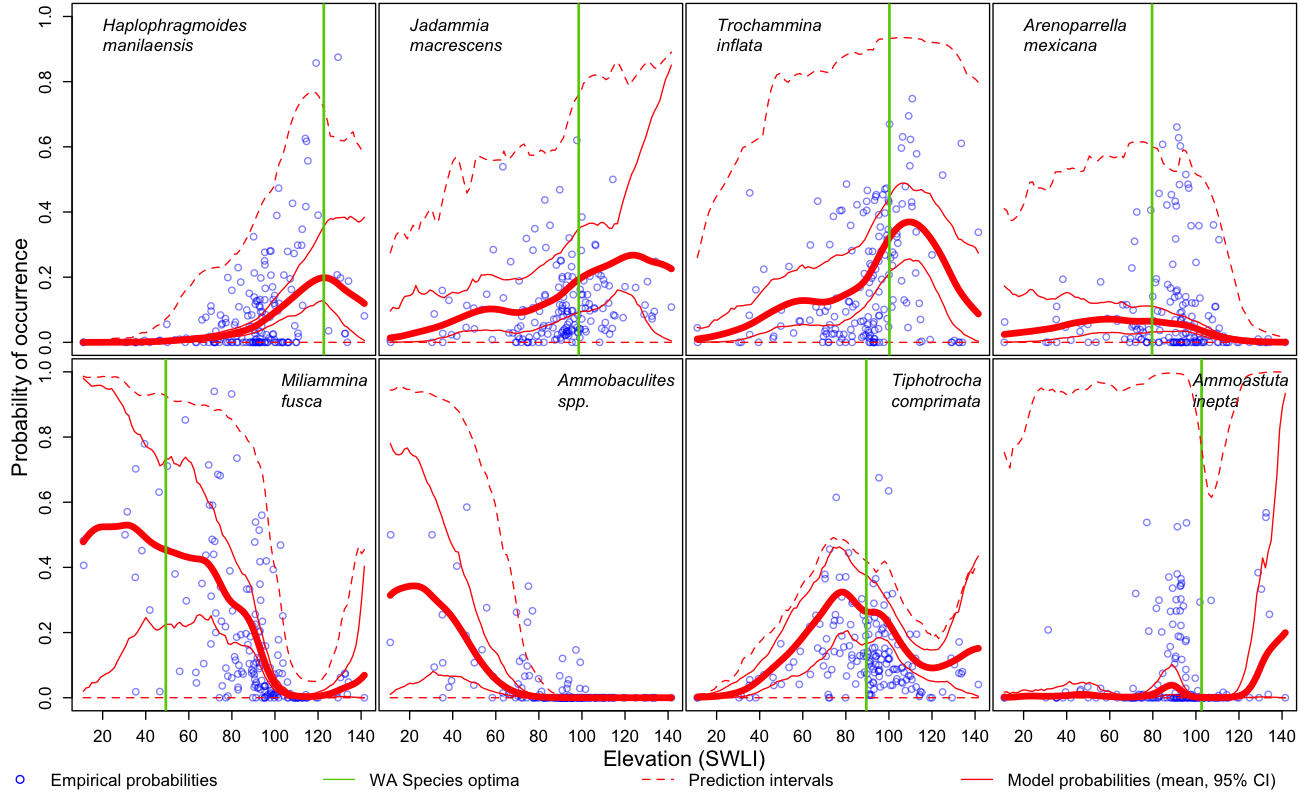}
\caption{The response of foraminifera species to elevation estimated from the modern training set using the Bayesian transfer function. The blue circles represent the probabilites of species occurance as determined from the raw count data (empirical probabilities). The response probabilites of occurance estimated by the Bayesian transfer function model are shown in red with a mean (heavy line), a credible interval for the mean (light line), and a prediction interval (dashed line). The green vertical lines represent the species optimum determined from the weighted average transfer function. }
\label{RCFig}
\end{center}
\end{figure}

The B-TF estimated a response curve (mean with a 95\% credible interval) for each species of foraminifera (expressed as raw counts) to tidal elevation (expressed as SWLI) from the modern training set of 172 samples (Figure \ref{RCFig}). The response curves are estimated from a multinomial distribution (in which species compositions are considered as a whole) parameterized by a probability vector p, which is the probability of a species being present at a given tidal elevation. The multinomial model (described in detail in Section \ref{multimodel}) utilises the combined species information from these observed responses to provide estimates for PME. The species prediction intervals (dashed red lines in Figure \ref{RCFig}) will aid in providing uncertainty for the PME estimates.\\

Broadly, we identify two forms of species-response curve in southern New Jersey.  First, a skewed, unimodal form describes the distribution of \emph{Haplophragmoides manilaenis} with a maximum probability of occurrence of $\sim$0.2 at 123 SWLI compared to a WA-TF species optimum that was also 123 SWLI.  \emph{Jadammia macrescens} and \emph{Trochammina inflata} also have a skewed, unimodal form with the highest probability of occurrences found in high salt-marsh environments at 124 SWLI (p $\approx$ 0.3) and 110 SWLI (p $\approx$ 0.4) respectively. The species optima for these species were situated at lower elevations by the WA-TF (99 SWLI and 100 SWLI respectively).  Both \emph{Ammobaculities spp.} (highest probability of occurrence of $\sim$0.3 at 22 SWLI) and \emph{Miliammina fusca} (highest probability of occurrence $\sim$0.5 at 31 SWLI) have skewed unimodal distributions with maximum probability of occurrence in low salt-marsh environments. The WA-TF estimated species optima of -19 and 49.33 SWLI respectively for these two species. \emph{Ammoastuta inepta} also has a skewed, unimodal form (maximum probability of occurrence of $\sim$ 0.2 at 140 SWLI). This species generally has a low probability of being present  because its distribution in southern New Jersey is restricted to sites with brackish salinity such as those located up river in Great Egg Harbor (Figure 1).  Relatively few samples from these environments are included in the modern training set and therefore in the dataset as a whole it is a rare, but ecologically-important, species.  Second, a unimodal Gaussian-like form describes the distributions of \emph{Arenoparrella mexicana} (maximum probability of occurrence of $\sim$0.07 between 60 and 70 SWLI) and Tiphotrocha comprimata (maximum probability of occurrence of $\sim$0.3 at 78 SWLI). In this case the WA-TF indicated species optima of 80 and 90 SWLI respectively for these two species. These results suggest that the number and type of ecological response curve prescribed to all species in the WA-TF model (and other transfer functions) might be inappropriate for accurately describing the relationship between salt-marsh foraminifera and tidal elevation.

\subsubsection{Cross validation of the modern data}
\label{CV}
\begin{figure}[h!]
\begin{center}
\includegraphics[scale=1.05]{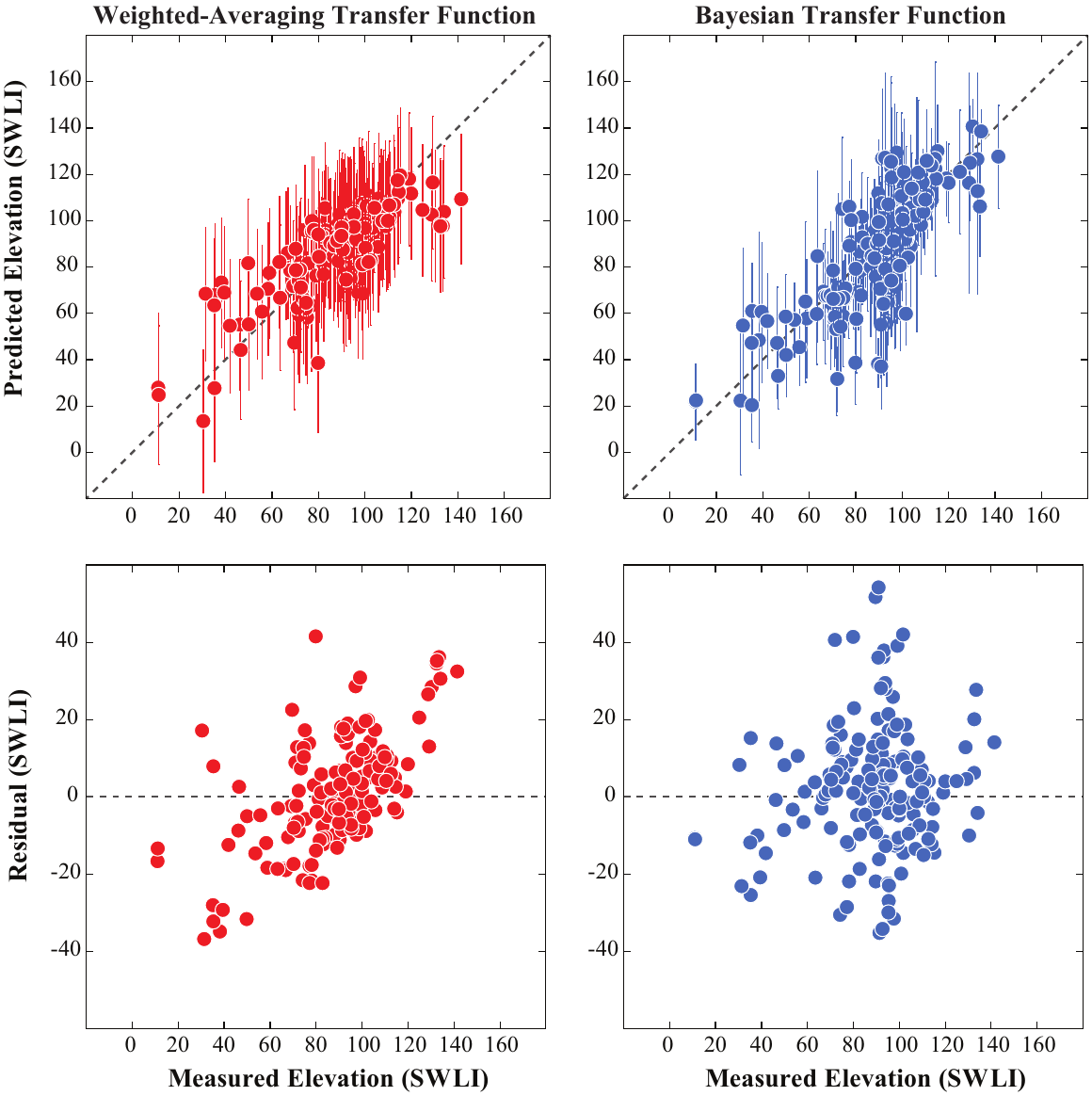}
\caption{Cross validation of the modern training set for the weighted average transfer function (red) and the Bayesian transfer function (blue). Upper panels are elevations in standardised water level index (SWLI) units, with lines representing $\pm$2$\sigma$ uncertainty for prediction. Lower panels show the (observed-predicted) residuals.}
\label{CVFig}
\end{center}
\end{figure}

Performance of the new B-TF and existing WA-TF was judged using 10-fold cross validation (Figure \ref{CVFig}).The uncertainty bounds ($\pm$2$\sigma$) for elevations predicted by the B-TF contained the true elevation 90\% of the time compared to 92\% for the WA-TF. The average 2$\sigma$ uncertainties are larger in the WA-TF (28 SWLI) than in the B-TF (21 SWLI).  The pattern of residuals in the WA-TF displayed a structure in which the elevation of low salt-marsh samples is over predicted (negative residuals) and the elevation of high salt-marsh samples is under predicted (positive residuals).  For example, the WA-TF showed an average residual of -16.6 between $\sim$10 and $\sim$70 SWLI and an average residual of 22.5 between $\sim$120 and $\sim$140 SWLI. This structure is absent in the B-TF suggesting that this model is better suited to reconstructing values close to the extremes of the sampled elevational gradient.

\newpage
\subsubsection{PME reconstructions}
\label{PME recon}
We reconstructed PME in the Cape May Courthouse and Leeds Point sediment cores using the WA-TF and B-TF models. At Cape May Courthouse, the B-TF estimated an average PME close to MHHW (SWLI=100) of 96.2 SWLI, with a standard deviation 14.1 SWLI (Figure \ref{ResultsFig}C, upper panel). The WA-TF also estimated an average PME close to MHHW of 96.7 SWLI, with a standard deviation of 4.1 SWLI (Figure \ref{ResultsFig}B, upper panel). The 2$\sigma$ uncertainties for the B-TF reconstructions ranged from 15.1 to 45.7 and are more variable than those from the WA-TF (28.1 to 29.0 SWLI). \\

At Leeds Point, the B-TF estimated an average PME close to MHHW of 92.8 SWLI, with a standard deviation 12.8 SWLI (Figure \ref{ResultsFig}C, lower panel). The WA-TF estimated PME close to MHHW  for all samples except for the 3.00-2.80 m interval where Miliammina fusca was present and PME reconstructions were correspondingly lower (Figure \ref{ResultsFig}B, lower panel). The 2$\sigma$ uncertainties for the B-TF reconstructions ranged from 11.5 to 59.8 SWLI and were more variable than those from the WA-TF (28.0 to 28.5 SWLI). \\

Comparison of the two models shows that the B-TF typically reconstructed PME with greater variability among samples than the WA-TF model. Similarly, reconstruction uncertainties were more variable for the B-TF model than the WA-TF model. Within their uncertainties the PME reconstructions for the B-TF and WA-TF overlap for all samples in both sediment cores.

\subsubsection{Multi-proxy reconstruction of PME}
\label{multi}
Measurements of $\delta^{13}$C in bulk-organic sediment are useful sea-level proxies because they readily distinguish between sediment that accumulated above MHHW in an environment dominated by C$_3$ plants and sediment that accumulated below MHHW in an environment dominated by C$_4$ plants.  This additional paleoenvironmental information was combined with the B-TF to generate a multi-proxy reconstruction of PME. The inclusion of the $\delta^{13}$C did not treat MHHW as a hard bound for PME, rather, if a sample is dominated by C$_3$ plants then the probability of PME being above MHHW increases. Likewise if a sample is dominated by C$_4$ plants the probability of PME being below MHHW water increases. \\

For Cape May Courthouse using the downcore $\delta^{13}$C values as a secondary proxy in the B-TF estimated an average PME of 87.5 SWLI (a reduction of 9 SWLI compared to the original B-TF). The 2$\sigma$ uncertainties estimated for the PME reconstructions were reduced by 32\% on average and up to $\sim$60\% for some samples (Figure \ref{ResultsFig}E, upper panel). For the Leeds Point core, the inclusion of the secondary $\delta^{13}$C proxy resulted in an estimated average PME of 86.1 SWLI (a reduction of 7 SWLI compared to the original B-TF). The 2$\sigma$ uncertainties decreased by $\sim$25\% on average  (Figure \ref{ResultsFig}E, lower panel). However, for samples where $\delta^{13}$C values and the presence of foraminifera indicate deposition in the transitional marsh (above MHHW, but below the highest occurrence of foraminifera) the uncertainty was reduced by an average of $\sim$50\% and up to $\sim$70\% for some samples because the constraint on PME changed from 0-150 SWLI to 100-150 SWLI.  These results demonstrate that incorporating a second line of proxy evidence in the B-TF framework is an efficient and formalized way to reduce uncertainty in RSL reconstructions in some sedimentary environments.

\subsection{Comparison among reconstructions}
\label {process}

\begin{figure}[h!]
\begin{center}
\includegraphics[scale=0.065]{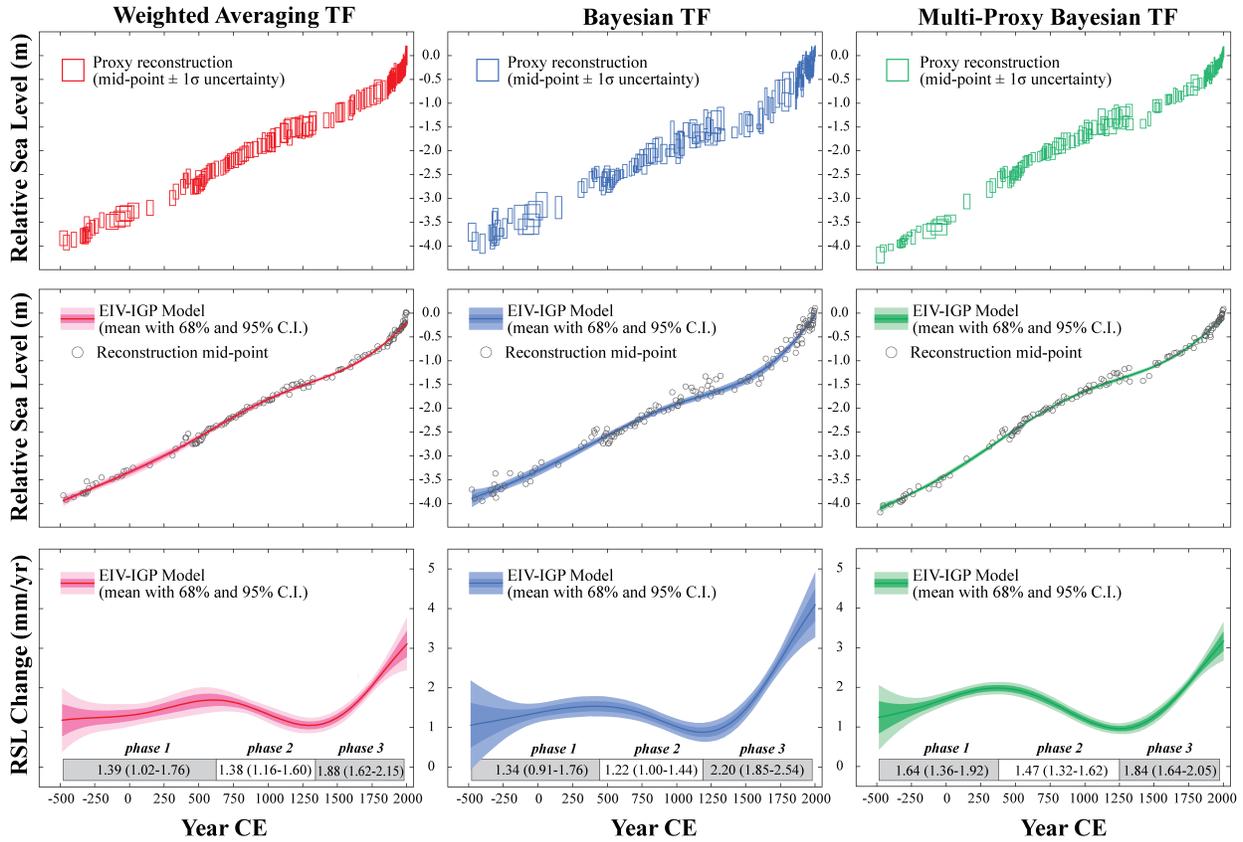}
\caption{The EIV-IGP model results for reconstructions produced using the B-TF, the WA-TF and the multi- proxy Bayesian transfer function. The upper panel shows individual data points (represented by rectangular boxes that illustrate the 95\% confidence region) and include age and relative sea-level uncertainties. The middle panels show the posterior fit of the errors-in-variables integrated Gaussian process model to the relative sea-level reconstructions.  Solid line represents the mean fit with the 68\% and 95\% credible intervals (C.I.) denoted by shading. The lower panels are the rates of relative sea-level (RSL) change. Shading denotes 68\% and 95\% credible intervals (C.I.) for the posterior mean of the rate process. The average rate for each phase of the reconstruction is given (in mm/yr) with a 95\% credible interval.}
\label{IGP}
\end{center}
\end{figure}

We applied the EIV-IGP model to the RSL reconstructions produced from the WA-TF, B-TF and multi-proxy B-TF to describe RSL trends along the coast of southern New Jersey since $\sim$500BCE (Figure \ref{IGP}). The WA-TF and B-TF models show $\sim$3.9 m of RSL rise compared to $\sim$4.1 m of RSL rise for the multi-proxy B-TF model. The multi-proxy B-TF reconstructed lower RSL at the beginning of the record ($\sim$ -4.2 m) compared to the B-TF and the WA-TF ($\sim$ -3.8 m) because of the additional constraint placed on the lowermost section of the Leeds Point core by $\delta^{13}$C values that indicate deposition at 100-150 SWLI.\\

All of the reconstructions show three phases of RSL behavior (Figure \ref{IGP}). The period from $\sim$500 BCE to $\sim$500 CE is a characterized by a continuous increase in the rate of RSL rise. The second stage shows a decline in rates of RSL rise from $\sim$500 CE to $\sim$1400 CE. After 1400 CE there is a continuous acceleration in the rate of RSL rise until reaching historic rates, which are unprecedented for at least 2500 years.\\

\begin{figure}[h!]
\begin{center}
\includegraphics[scale=0.8]{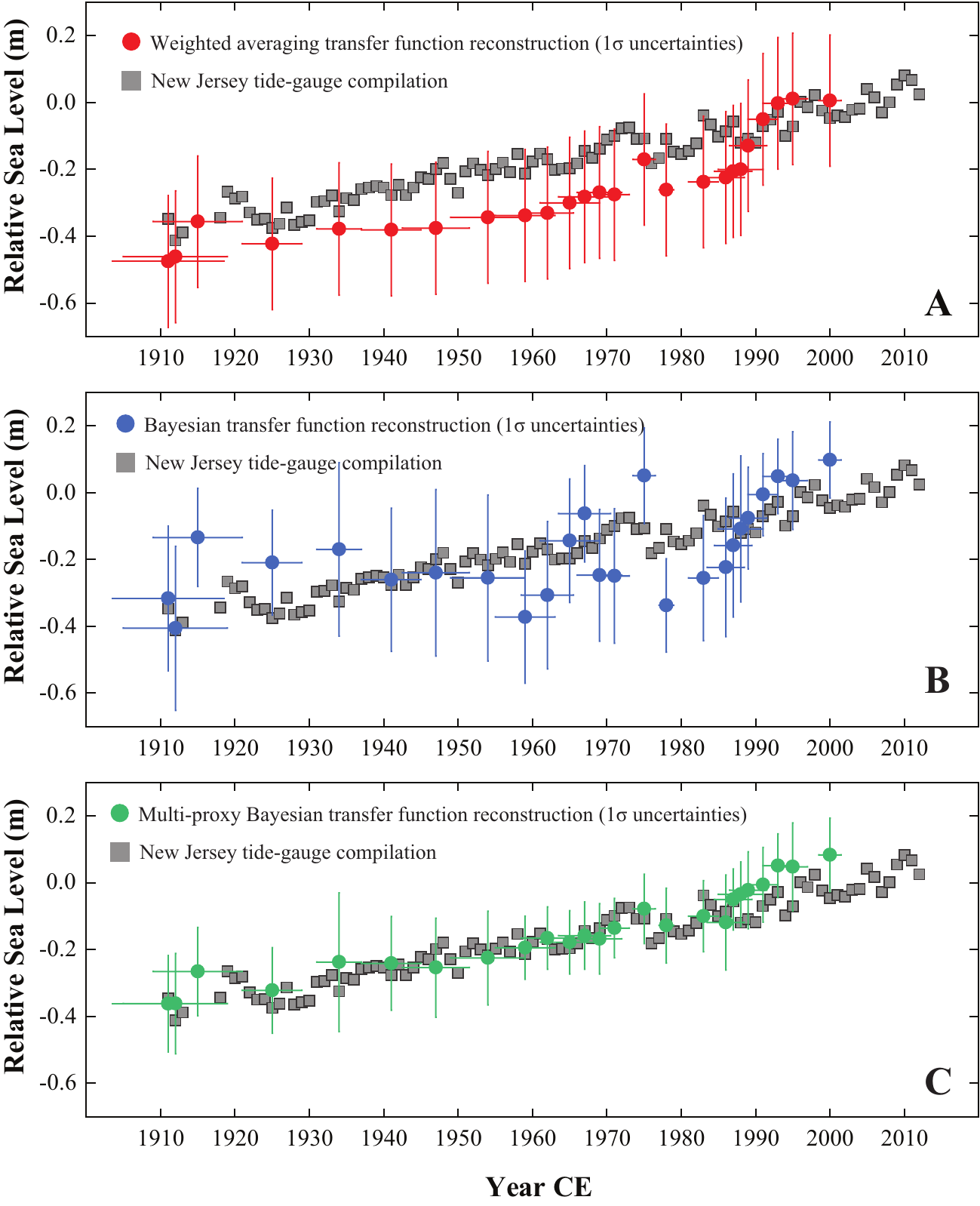}
\caption{Comparison of the weighted average transfer function (A), the Bayesian transfer function (B) and the multi-proxy Bayesian transfer function (C) relative sea-level reconstruction with tide gauge data observed in the New Jersey region. The lines represent $\pm$1$\sigma$ uncertainty for the reconstruction. }
\label{TGcompare}
\end{center}
\end{figure}

\newpage
However, there are some differences among the three reconstructions. For example, the B-TF shows the highest modern rate of rise at 4.1 mm/yr (95\% C.I. 3.27-4.92 mm/yr) in 2000 CE compared to 3.16 mm/yr (95\% C.I. 2.68-3.65 mm/yr) and 3.11 mm/yr (95\% C.I. 2.45-3.77 mm/yr) for the multi-proxy B-TF and the WA-TF respectively. The B-TF consistently estimated RSL lower than the multi-proxy B-TF and the WA-TF between $\sim$1400 to $\sim$1700, resulting in the observed difference in the rates into the 21st century. When compared to the observed tide-gauge data for the last $\sim$100 years from New Jersey (Figure \ref{TGcompare}), the quality of the estimated RSL mid-point reconstructions can be assessed using mean squared error (MSE). For the multi-proxy B-TF the MSE was estimated at 0.003 m$^2$ compared to 0.014 m$^2$ for the B-TF and the WA-TF, indicating that the multi-proxy B-TF mid-points provide better estimates for RSL in comparison to the B-TF and the WA-TF.

\section{Discussion}
The B-TF provides an alternative to the (non-Bayesian) regression-based transfer functions commonly used for reconstructing RSL (e.g., \citealp{Horton1999, Gehrels2000, Barlow2014}) and in conjunction with the previously developed chronology and process modules enables RSL to be reconstructed in an entirely Bayesian framework. A key difference between the B-TF and existing transfer functions (e.g. the WA-TF) is the modeled relationship between species of foraminifera and tidal elevation. The number and type of species-response curves estimated by the B-TF model stands in contrast to the WA-TF that assumes a unimodal Gaussian form for all species. The optima and tolerance estimated for each species by the WA-TF shows overlap with the B-TF species-response curves, particularly those that have a Gaussian form such as \emph{Tiphotrocha comprimata}. However, this form is only appropriate for two of the eight dominant species in the southern New Jersey training set. The flexible species-specific response provided by the B-TF is more appropriate given that models based on the assumption of a single response do not adequately explain the ecological behavior of the dominant species in New Jersey, or other salt marsh foraminiferal assemblages (e.g., \citealp{EdwardsandHorton2006}), or species from other biological groups used in RSL reconstructions such as diatoms (e.g., \citealp{Zong1999, Gehrels2000}). This variability in species response to a single environmental variable arises from ecological complexity and the influence of secondary environmental variables (e.g., \citealp{Murray1991}). Therefore, we propose that the additional flexibility of the B-TF will produce more accurate PME (and consequently RSL) reconstructions than existing transfer functions such as the WA-TF.\\

The implication of the flexibility of the B-TF is illustrated in the cross validation results. The WA-TF appeared to suffer from ‘edge effects’ (i.e., a tendency for the model to bias PME predictions towards the mean of the training data), a common artifact of using weighted average based methods \citep{ter1993,Birks95}. Our B-TF does not suffer from this prediction bias and outperformed the WA-TF in the upper and lower extremes of tidal elevation. The consequences of such an improvement are significant where true PMEs lie close to the ends of the sampled environmental gradient. For example, on subduction zone coastlines such as the Pacific Northwest coast of North America (e.g., \citealp{Nelson1996}, cyclical tectonic activity contributes to reconstructed RSL trends. During a slow (100s to 1000s of years) inter-seismic phase, accumulation of strain results in uplift of the coast (RSL fall). Conversely, the strain is released during an instantaneous co-seismic phase in which the coastline subsides (RSL rise). These processes cause significant and very rapid shifts in depositional environment that can span the full elevational range of coastal environments from sub-tidal settings to supra-tidal, freshwater uplands. In contrast, the sediment sequences targeted for reconstructing Common Era RSL on passive margins (e.g. New Jersey) are commonly comprised of unbroken sequences of high salt-marsh peat that are less susceptible to these edge effects.\\

Further motivation for the development of a Bayesian model for RSL reconstruction lies in the quantification of reconstruction uncertainty. Non-Bayesian transfer function methods (e.g. the WA-TF model) assume that model parameters are fixed and known. Therefore, they do not incorporate uncertainty into the estimation of the PME reconstruction itself, rather, the uncertainty is produced separately either before or after PME was estimated. This uncertainty is the root mean square error from two sources (S1 and S2; \citealp{Birks1990, JugginsandBirks2012}). The S1 contribution is sample-specific and is the standard deviation of bootstrapped PME reconstructions. The S2 contribution is the difference between observed and predicted tidal elevations established by cross validation of the modern training set (Figure \ref{CVFig}). The uncertainties for PME reconstructed by the WA-TF model show very little variability among samples (2$\sigma$ ranges from 28.0 to 29.0). This pattern arises because the contribution from the sample-specific (S1) uncertainty is very small compared to the model uncertainty (S2) which is common to all samples. As a result the PME reconstructions for all sediment core samples have very similar uncertainties despite biological variability in species composition.\\

Alternatively, Bayesian methods explicitly model the uncertainty associated with individual reconstructions. Uncertainty for PME (and other unknown parameters) is included in the probability model through prior distributions. Assuming distributions for unknown parameters (in contrast to non-Bayesian approaches that use point estimates) allows the parameter uncertainty from the calibration step to be formally propagated into the reconstruction step. Therefore, estimates of PME produced by the B-TF take fuller account of the uncertainties related to the model and its parameters than non-Bayesian counterparts. The uncertainties estimated by the B-TF (excluding a secondary proxy) show more pronounced variability among core samples (2$\sigma$ uncertainties range between 15.1 to 45.7). This variability arises from the observed response distribution of each species to tidal elevation (estimated from the modern data; Figure \ref{RCFig}). For each individual species there is variability in both the uncertainty of the mean response curves and in the prediction intervals (i.e. uncertainty is greater in some parts of the elevational gradient than at others).\\

The variability of reconstructed PME from the B-TF is a more accurate reflection of the observed trends in downcore foraminiferal populations and is therefore a more ecologically plausible reconstruction than the WA-TF model. In the New Jersey case study in both cores the dominant groups of foraminifera are characteristic of a high salt-marsh environment. \citet{Engelhart2012} concluded that samples identified as being of high salt marsh origin formed between MHW (SWLI values of 93 at Cape May and 90 at Leeds Point) and HAT (SWLI values of 123 at Cape May and 127 at Leeds Point). But there is a pronounced lack of variability reconstructed PME using the WA-TF model (average of $\sim$95 SWLI with a standard deviation of 5.5). This lack of variability in reconstructed PME is at odds with the observed downcore variability in species assemblages. For example, the key, high salt-marsh species \emph{Jadammina macrescens} and \emph{Trochammina inflata} vary in relative abundance from 0\% (absent) to 100\% and approximately 80\%, respectively (Figure 3). In contrast, PME reconstructions from the B-TF are also estimated at an average of $\sim$95 SWLI, but with a larger standard deviation of 13.1.\\

The majority of quantitative RSL reconstructions employ a single proxy (e.g., \citealp{Kemp2011}). A number of other proxies are available to support RSL reconstructions primarily produced from salt-marsh foraminifera. Additional biological proxies could include different groups of organisms with a relationship to tidal elevation such as diatoms (e.g., \citealp{Zong1999,Shennan94}) or thecomebians (e.g., \citealp{Charman2010, Roe2002}). These organisms can be incorporated as presence/absence data or as species counts from a modern training set of paired observations of species abundance and tidal elevation. A number of lithological proxies (e.g., \citealp{Nelson2015}) are also available which can be qualitative (such as field and lab-based descriptions of sediment as high marsh or low marsh) or quantitative (such as measurements of organic content; e.g., \citealp{Plater2015}) and may provide thresholds in a similar fashion to sediment geochemistry in New Jersey. Although secondary proxies are often available to provide additional and independent constraints, a barrier to their use is the lack of an accessible and formal framework for combining multiple proxies with appropriate consideration of uncertainty. A strength of our B-TF is its ability to accommodate these secondary proxy sources. In the example from New Jersey we primarily used a biological proxy (assemblages of foraminifera), but amended the model to include information from a geochemical proxy (bulk sediment $\delta^{13}$C). On average this approach reduced the uncertainty for PME reconstructions by $\sim$28\%. The reduction in uncertainty consequently provides constraints on this history of RSL change and more precise estimates of rates of RSL change through time. This is highlighted in the reconstruction of sea level between $\sim$ 500 BCE and 500 CE where the multi-proxy B-TF shows rates of rise for this period reach a maximum of $\sim$1.9 mm/yr which is greater than the rates produced by the B-TF and the WA-TF ($\sim$1.5 and 1.6 mm/yr respectively). Uncertainty for these rate estimates was reduced by 25\% for the multi-proxy B-TF compared to the WA-TF and the B-TF. These results highlight the specific utility of bulk sediment $\delta^{13}$C measurements as a sea-level indicator along the Atlantic coast of North America and the general utility of employing a multi-proxy approach to reconstructing RSL where the goal is to produce reconstructions with the best possible precision.\\

A practical and intuitive means to illustrate the improved performance of the B-TF over the WA-TF model is to compare RSL reconstructions with long-term measurements made by nearby tide gauges (\citealp{Kemp2009,Kemp13,Gehrels2000,Barlow2014,Long2014,Leorri2008}). We compare the reconstruction provided by the WA-TF, B-TF and multi-proxy B-TF with regional tide-gauge measurements from New Jersey (Figure \ref{TGcompare}). The tide-gauges measured about 30 cm of RSL change over the period 1911 to 2012 CE. Considering the 1$\sigma$ errors in the reconstructions are of the order $\pm$10 cm it is unsurprising that the uncertainty bounds of reconstructions capture the tide gauge observations. However, the RSL mid-points reconstructed by the multi-proxy B-TF are notably better at capturing the variability observed in the tide-gauge data. This suggests that the variability in the PME estimates produced from B-TF is relevant (the model is picking up a signal (as opposed to noise) due to downcore changes in foraminifera assemblages) and the estimates are improved by the addition of a secondary proxy. The improved fit between the instrumental records and the proxy reconstruction using the multi-proxy B-TF model indicates that it is possible to reconstruct decadal to multi-decadal RSL trends using salt-marsh sediment in New Jersey and similar regions.  This is a noticeable advantage over existing approaches such as the WA-TF that reconstruct multi-decadal to centennial scale trends because in the absence of reconstructed PME variability, the resulting RSL reconstructions are primarily driven by the age-depth model.

\section{Conclusion}
To accurately reconstruct the continuous and dynamic evolution of sea-level change, we developed a Bayesian hierarchical model comprised of three formally interconnected modules. (1) A B-TF for the calibration of foraminifera into tidal elevation, which is flexible enough to formally accommodate additional proxies (bulk-sediment $\delta^{13}$C). (2) An existing chronology developed from a Bchron age-depth model. (3) An existing EIV-IGP model for estimating rates of sea-level change.  Previous reconstructions treated these three components as independent and employed existing approaches that were developed in a variety of numerical frameworks.  \\

Our new B-TF provides an alternative to existing transfer functions. We illustrate the improved performance of our approach by applying the B-TF and a WA-TF model to a dataset of common Era salt-marsh foraminifera from southern New Jersey, U.S.A. The relationship between species of salt marsh foraminifera and tidal elevation was described using a regional-scale modern training set (n = 172) comprised of paired observations of species abundance and elevation. Results from the B-TF show that six of the eight most dominant foraminifera do not conform to the unimodal, Gaussian response curve prescribed by the WA-TF and other existing transfer functions.\\

We propose that the B-TF produces more accurate RSL reconstructions with a more complete evaluation of uncertainty and greater ecological plausibility than the WA-TF model. We applied the transfer functions to cores of salt-marsh sediment that were recovered from two sites in southern New Jersey. The flexible approach utilized in the B-TF results in more variability in reconstructed PME and associated uncertainty  among samples than the WA-TF model. This variability is consistent with observed changes in foraminiferal population in core samples. \\

The B-TF allows results from additional, independent sea-level proxies to be formally incorporated alongside the primary biological proxy to produce a multi-proxy reconstruction. In New Jersey, we used bulk sediment $\delta^{13}$C values to determine if a core sample formed above or below the MHHW tidal datum. The addition of a second proxy reduced reconstruction uncertainty by an average of 28\% and up to $\sim$70\% for some samples. \\

We assess the ability of the multi-proxy B-TF, B-TF and the WA-TF to reconstruct RSL through comparison with observed tide-gauge data from New Jersey. Results showed that the 2$\sigma$ uncertainty bounds for all reconstructions capture the observations from the tide gauge. However, the multi-proxy B-TF provides improved estimates (MSE = 0.003 m$^2$) for the reconstructed RSL mid points compared to the B-TF and the WA-TF (MSE = 0.014 m$^2$), indicating that the multi-proxy B-TF has the potential to capture the decadal-scale variability seen in the tide gauge data.

\section*{Acknowledgements}
Supported by the Structured PhD in Simulation Science which is funded by the Programme for Research in Third Level Institutions (PRTLI) Cycle 5 and co-funded by the European Regional Development Fund, and the Science Foundation Ireland Research Frontiers Programme (2007/RFP/MATF281)  and also supported by the National Science Foundation awards EAR and OCE 1458904.

\bibliographystyle{plainnat}
\bibliography{/Users/niamhcahill/Documents/BibTex/TF.bib}

\begin{thebibliography}{66}
\providecommand{\natexlab}[1]{#1}
\providecommand{\url}[1]{\texttt{#1}}
\expandafter\ifx\csname urlstyle\endcsname\relax
  \providecommand{\doi}[1]{doi: #1}\else
  \providecommand{\doi}{doi: \begingroup \urlstyle{rm}\Url}\fi

\bibitem[Barlow et~al.(2014)Barlow, Long, Saher, Gehrels, Garnett, and
  Scaife]{Barlow2014}
N~L.~M. Barlow, A~J. Long, M~H. Saher, W~R. Gehrels, M~H. Garnett, and R~G.
  Scaife.
\newblock Salt-marsh reconstructions of relative sea-level change in the
  {N}orth {A}tlantic during the last 2000 years.
\newblock \emph{Quaternary Science Reviews}, 99:\penalty0 1--16, 2014.

\bibitem[Birks(1995)]{Birks95}
H~J.~B. Birks.
\newblock Quantitative {P}alaeoenvironmental {R}econstructions.
\newblock In D~Maddy and J~S. Brew, editors, \emph{{S}tatistical {M}odelling of
  {Q}uaternary {S}cience {D}ata}, volume Technical Guide 5 of \emph{Technical
  Guide}, pages 161--254. Quaternary Research Association, Cambridge, 1995.

\bibitem[Birks et~al.(1990)Birks, Line, Juggins, Stevenson, and ter
  Braak]{Birks1990}
H~J.~B. Birks, J~M. Line, S~Juggins, A~C. Stevenson, and C~J.~F. ter Braak.
\newblock Diatoms and p{H} reconstruction.
\newblock \emph{Philosophical Transactions of the Royal Society of London B:
  Biological Sciences}, 327\penalty0 (1240):\penalty0 263--278, 1990.

\bibitem[Birks(2012)]{Birks2012}
H~J.B. Birks.
\newblock Overview of numerical methods in palaeolimnology.
\newblock In H.J.B. Birks, A.F. Lotter, S.~Juggins, and J.P. Smol, editors,
  \emph{{T}racking {E}nvironmental {C}hange {U}sing {L}ake {S}ediments: {D}ata
  {H}andling and {N}umerical {T}echniques}, volume~5 of \emph{Tracking
  environmental change using lake sediments}, book section~2. Springer, 2012.

\bibitem[Birks(2010)]{PAGES}
J~Birks.
\newblock Calibration, {T}ransfer {F}unctions and {E}nvironmental
  {R}econstructions.
\newblock In \emph{PAGES}, October 2010.

\bibitem[Blaauw and Christen(2011)]{blaauw2011}
M~Blaauw and J~A. Christen.
\newblock Flexible paleoclimate age-depth models using an autoregressive gamma
  process.
\newblock \emph{Bayesian Anal.}, \penalty0 (3):\penalty0 457--474, 09 2011.

\bibitem[Bradley(2015)]{Bradley2015}
R~S. Bradley.
\newblock Chapter 1 - {P}aleoclimatic {R}econstruction.
\newblock In R~S. Bradley, editor, \emph{Paleoclimatology (Third Edition)},
  pages 1--11. Academic Press, San Diego, third edition edition, 2015.

\bibitem[Bronk~Ramsey(2008)]{OxCal}
C~Bronk~Ramsey.
\newblock Radiocarbon {D}ating: {R}evolutions in {U}nderstanding*.
\newblock \emph{Archaeometry}, 50\penalty0 (2):\penalty0 249--275, 2008.

\bibitem[Cahill et~al.(2015)Cahill, Kemp, Horton, and Parnell]{Cahill15}
N~Cahill, A~C. Kemp, B~P. Horton, and A~C. Parnell.
\newblock Modeling sea-level change using errors-in-variables integrated
  gaussian processes.
\newblock \emph{Annals of Applied Statistics}, 9\penalty0 (2):\penalty0
  547--571, 2015.

\bibitem[Charman et~al.(2010)Charman, Gehrels, Manning, and
  Sharma]{Charman2010}
D~J. Charman, W~R. Gehrels, C~Manning, and C~Sharma.
\newblock Reconstruction of recent sea-level change using testate amoebae.
\newblock \emph{Quaternary Research}, 73\penalty0 (2):\penalty0 208--219, 2010.

\bibitem[De~Boor(1978)]{deboor78}
C~De~Boor.
\newblock \emph{{A {P}ractical {G}uide to {S}plines}}.
\newblock Springer, 1978.

\bibitem[de~Rijk(1995)]{deRijk1995}
S~de~Rijk.
\newblock Salinity control on the distribution of salt marsh foraminifera
  ({G}reat {M}arshes, {M}assachusetts).
\newblock \emph{The Journal of Foraminiferal Research}, 25\penalty0
  (2):\penalty0 156--166, 1995.

\bibitem[Dey et~al.(2000)Dey, Ghosh, and Mallick]{EIV}
D~K. Dey, S~K. Ghosh, and B~K. Mallick.
\newblock \emph{{G}eneralized {L}inear {M}odels {A} {B}ayesian {P}erspective}.
\newblock Marcel Dekker, Inc, 2000.

\bibitem[Dierckx(1993)]{Dierckx93}
P~Dierckx.
\newblock \emph{{Curve and Surface Fitting with Splines}}.
\newblock Clarendon Press, Oxford, 1993.

\bibitem[Douglas(1991)]{Douglas91}
B~C. Douglas.
\newblock Global sea level rise.
\newblock \emph{Journal of Geophysical Research: Oceans}, 96\penalty0
  (C4):\penalty0 6981--6992, 1991.

\bibitem[Dutton et~al.(2015)Dutton, Carlson, Long, Milne, Clark, DeConto,
  Horton, Rahmstorf, and Raymo]{Dutton2015}
A.~Dutton, A.~E. Carlson, A.~J. Long, G.~A. Milne, P.~U. Clark, R.~DeConto,
  B.~P. Horton, S.~Rahmstorf, and M.~E. Raymo.
\newblock Sea-level rise due to polar ice-sheet mass loss during past warm
  periods.
\newblock \emph{Science}, 349\penalty0 (6244), 2015.

\bibitem[Edwards and Horton(2006)]{EdwardsandHorton2006}
R~J. Edwards and B~P. Horton.
\newblock Developing detailed records of relative sea-level change using a
  foraminiferal transfer function: an example from {N}orth {N}orfolk, {UK}.
\newblock \emph{Philosophical Transactions of the Royal Society A -
  Mathematical Physical and Engineering Sciences}, 364\penalty0
  (1841):\penalty0 973--991, 2006.

\bibitem[Edwards et~al.(2004)Edwards, Wright, and van~de Plassche]{Edwards2004}
R~J. Edwards, A~J. Wright, and O~van~de Plassche.
\newblock Surface distributions of salt-marsh foraminifera from {C}onnecticut,
  {USA}: modern analogues for high-resolution sea level studies.
\newblock \emph{Marine Micropaleontology}, 51:\penalty0 1--21, 2004.

\bibitem[Edwards and Wright(2015)]{EdwardsandWright}
R.J. Edwards and A.~J. Wright.
\newblock Foraminifera.
\newblock In I.~Shennan, A.~J. Long, and B.~P. Horton, editors,
  \emph{{H}andbook of {S}ea-{L}evel {R}esearch}, book section~13, pages
  191--217. John Wiley \& Sons, 2015.

\bibitem[Engelhart and Horton(2012)]{Engelhart2012}
S~E. Engelhart and B~P. Horton.
\newblock Holocene sea level database for the {A}tlantic coast of the {U}nited
  {S}tates.
\newblock \emph{Quaternary Science Reviews}, 54:\penalty0 12--25, 2012.

\bibitem[Fritz et~al.(1991)Fritz, Juggins, Battarbee, and Engstrom]{Fritz1991}
S.~C. Fritz, S.~Juggins, R.~W. Battarbee, and D~Engstrom.
\newblock Reconstruction of past changes in salinity and climate using a
  diatom-based transfer function.
\newblock \emph{Nature}, 352\penalty0 (6337):\penalty0 706--708, 1991.

\bibitem[Gehrels(1994)]{Gehrels1994}
W.~R Gehrels.
\newblock Determining relative sea-level change from salt-marsh foraminifera
  and plant zones on the coast of {M}aine, {U.S.A.}
\newblock \emph{Journal of Coastal Research}, 10\penalty0 (4):\penalty0
  990--1009, 1994.

\bibitem[Gehrels(2000)]{Gehrels2000}
W.~R Gehrels.
\newblock Using foraminiferal transfer functions to produce high-resolution
  sea-level records from salt-marsh deposits, {M}aine, {USA}.
\newblock \emph{The Holocene}, 10\penalty0 (3):\penalty0 367--376, 2000.

\bibitem[Haslett and Parnell(2008)]{bchron}
J~Haslett and A~C. Parnell.
\newblock A simple monotone process with application to radiocarbon-dated depth
  chronologies.
\newblock \emph{Journal of the Royal Statistical Society: Series C (Applied
  Statistics)}, 57\penalty0 (4):\penalty0 399--418, 2008.

\bibitem[Haslett et~al.(2006)Haslett, Whiley, Bhattacharya, Salter-Townshend,
  Wilson, Allen, Huntley, and Mitchell]{Haslett2006}
J.~Haslett, M.~Whiley, S.~Bhattacharya, M.~Salter-Townshend, Simon~P. Wilson,
  J.~R.~M. Allen, B.~Huntley, and F.~J.~G. Mitchell.
\newblock Bayesian palaeoclimate reconstruction.
\newblock \emph{Journal of the Royal Statistical Society: Series A (Statistics
  in Society)}, 169\penalty0 (3):\penalty0 395--438, 2006.

\bibitem[Holsclaw et~al.(2013)Holsclaw, Sanso, Lee, Heitmann, Habib, Higdon,
  and Alam]{Holsclaw2013}
T~Holsclaw, B~Sanso, H~K.~H. Lee, K~Heitmann, S~Habib, D~Higdon, and Ui~Alam.
\newblock Gaussian process modeling of derivative curves.
\newblock \emph{Technometrics}, 55\penalty0 (1):\penalty0 57--67, 2013.

\bibitem[Horton and Edwards(2006)]{Horton06}
B~P. Horton and R~J. Edwards.
\newblock Quantifying {H}olocene sea-level change using intertidal
  foraminifera: lessons from the {B}ritish {I}sles.
\newblock \emph{Cushman Foundation for Foraminiferal Research, Special
  Publication}, 40:\penalty0 97, 2006.

\bibitem[Horton et~al.(1999)Horton, Edwards, and Lloyd]{Horton1999}
B~P. Horton, R~J. Edwards, and J~M. Lloyd.
\newblock {UK} intertidal foraminiferal distributions: implications for
  sea-level studies.
\newblock \emph{Marine Micropaleontology}, 36\penalty0 (4):\penalty0 205--223,
  1999.

\bibitem[Imbrie and Kipp(1971)]{Imbrie1971}
J~Imbrie and N~G. Kipp.
\newblock A new micropalaeontological method for quantitative paleoclimatology:
  application to a late {P}leistocene {C}aribbean core.
\newblock In K~K. Turekian, editor, \emph{The {L}ate {C}enozoic {G}lacial
  {A}ges}, pages 71--181. Yale University Press, New Haven and London, 1971.

\bibitem[Jackson and Williams(2004)]{Jackson&WIlliams04}
S.~T. Jackson and J.~W. Williams.
\newblock Modern analogs in {Q}uaternary paleoecology: {H}ere today, gone
  yesterday, gone tomorrow?
\newblock \emph{Annual Review of Earth and Planetary Sciences}, 32:\penalty0
  495--537, 2004.

\bibitem[Juggins and Birks(2012)]{JugginsandBirks2012}
S~Juggins and H~J.B. Birks.
\newblock {Q}uantiative {E}nvironmental {R}econstructions {F}rom {B}iological
  {D}ata.
\newblock In H.J.B. Birks, A.F. Lotter, S.~Juggins, and J.P. Smol, editors,
  \emph{{T}racking {E}nvironmental {C}hange {U}sing {L}ake {S}ediments: {D}ata
  {H}andling and {N}umerical {T}echniques}, volume~5 of \emph{Tracking
  environmental change using lake sediments}, pages 431--494. Springer, 2012.

\bibitem[Kemp et~al.(2009{\natexlab{a}})Kemp, Horton, Corbett, Culver, Edwards,
  and van~de Plassche]{Kemp09}
A~C. Kemp, B~P. Horton, D~R. Corbett, S~J. Culver, R~J. Edwards, and O~van~de
  Plassche.
\newblock The relative utility of foraminifera and diatoms for reconstructing
  late {H}olocene sea-level change in {N}orth {C}arolina, {USA}.
\newblock \emph{Quaternary Research}, 71\penalty0 (1):\penalty0 9--21,
  2009{\natexlab{a}}.

\bibitem[Kemp et~al.(2009{\natexlab{b}})Kemp, Horton, Culver, Corbett, van~de
  Plassche, Gehrels, Douglas, and Parnell]{Kemp2009}
A~C. Kemp, B~P. Horton, S~J. Culver, D.~R Corbett, O~van~de Plassche, W.~R
  Gehrels, B~C. Douglas, and A~C. Parnell.
\newblock Timing and magnitude of recent accelerated sea-level rise ({N}orth
  {C}arolina, {U}nited {S}tates).
\newblock \emph{Geology}, 37\penalty0 (11):\penalty0 1035--1038,
  2009{\natexlab{b}}.

\bibitem[Kemp et~al.(2011)Kemp, Horton, Donnelly, Mann, Vermeer, and
  Rahmstorf]{Kemp2011}
A~C. Kemp, B~P. Horton, J~P. Donnelly, M~E. Mann, M~Vermeer, and S~Rahmstorf.
\newblock Climate related sea-level variations over the past two millennia.
\newblock \emph{Proceedings of the National Academy of Sciences}, 108\penalty0
  (27):\penalty0 11017--11022, 2011.

\bibitem[Kemp et~al.(2012)Kemp, Vane, Horton, Engelhart, and
  Nikitina]{Kemp2012}
A~C. Kemp, C~H. Vane, B~P. Horton, S~E. Engelhart, and D~Nikitina.
\newblock Application of stable carbon isotopes for reconstructing salt-marsh
  floral zones and relative sea level, {N}ew {J}ersey, {USA}.
\newblock \emph{Journal of Quaternary Science}, 27\penalty0 (4), 2012.

\bibitem[Kemp et~al.(2013{\natexlab{a}})Kemp, Horton, Vane, Corbett, Bernhardt,
  Engelhart, Anisfeld, Parnell, and Cahill]{Kemp2013}
A~C. Kemp, B~P. Horton, C~H. Vane, D~R. Corbett, C~E. Bernhardt, S~E.
  Engelhart, S~C. Anisfeld, A~C. Parnell, and N~Cahill.
\newblock Sea-level change during the last 2500 years in {N}ew {J}ersey, {USA}.
\newblock \emph{Quaternary Science Reviews}, 81:\penalty0 90--104,
  2013{\natexlab{a}}.

\bibitem[Kemp et~al.(2013{\natexlab{b}})Kemp, Telford, Horton, Anisfeld, and
  Sommerfield]{Kemp13}
A~C. Kemp, R~J. Telford, B~P. Horton, S~C. Anisfeld, and C~K. Sommerfield.
\newblock Reconstructing {H}olocene sea-level using salt-marsh foraminifera and
  transfer functions: lessons from {N}ew {J}ersey, {USA}.
\newblock \emph{Journal of Quaternary Science}, 28\penalty0 (6):\penalty0
  617--629, 2013{\natexlab{b}}.

\bibitem[Korsman and Birks(1996)]{Korsman96}
T~Korsman and H~J.B. Birks.
\newblock Diatom-based water chemistry reconstructions from northern {S}weden:
  a comparison of reconstruction techniques.
\newblock \emph{Journal of Paleolimnology}, 15\penalty0 (1):\penalty0 65--77,
  1996.

\bibitem[Leorri et~al.(2008)Leorri, Horton, and Cearreta]{Leorri2008}
E~Leorri, B~P. Horton, and A~Cearreta.
\newblock Development of a foraminifera-based transfer function in the {B}asque
  marshes, {N}. {S}pain: Implications for sea-level studies in the {B}ay of
  {B}iscay.
\newblock \emph{Marine Geology}, 251\penalty0 (1-2):\penalty0 60--74, 2008.

\bibitem[Li et~al.(2010)Li, Nychka, and Ammann]{Bo2010}
B~Li, D~W. Nychka, and C~M. Ammann.
\newblock The value of multiproxy reconstruction of past climate.
\newblock \emph{Journal of the American Statistical Association}, 105\penalty0
  (491):\penalty0 883--895, 2010.

\bibitem[Long et~al.(2014)Long, Barlow, Gehrels, Saher, Woodworth, Scaife,
  Brain, and Cahill]{Long2014}
A~J. Long, N~L.~M. Barlow, W~R. Gehrels, M~H. Saher, P~L. Woodworth, R~G.
  Scaife, M~J. Brain, and N~Cahill.
\newblock Contrasting records of sea-level change in the eastern and western
  {N}orth {A}tlantic during the last 300 years.
\newblock \emph{Earth and Planetary Science Letters}, 388:\penalty0 110--122,
  2014.

\bibitem[Mann et~al.(2009)Mann, Zhang, Rutherford, Bradley, Hughes, Shindell,
  Ammann, Faluvegi, and Ni]{Mann2009}
M~E. Mann, Z~Zhang, S~Rutherford, R~S. Bradley, M~K. Hughes, D~Shindell,
  C~Ammann, G~Faluvegi, and F~Ni.
\newblock Global signatures and dynamical origins of the {L}ittle {I}ce {A}ge
  and {M}edieval {C}limate {A}nomaly.
\newblock \emph{Science}, 326\penalty0 (5957):\penalty0 1256--1260, 2009.

\bibitem[Murray(1991)]{Murray1991}
J~W. Murray.
\newblock \emph{{Ecology and {P}alaeoecology of {B}enthic {F}oraminifera}}.
\newblock Elsevier, Amsterdam, 1991.

\bibitem[Nelson(2015)]{Nelson2015}
A~R. Nelson.
\newblock Coastal sediments.
\newblock In I.~Shennan, A.~J. Long, and B.~P. Horton, editors,
  \emph{{H}andbook of {S}ea-Level {R}esearch}, pages 47--65. John Wiley \&
  Sons, 2015.

\bibitem[Nelson et~al.(1996)Nelson, Jennings, and Kashima]{Nelson1996}
A~R. Nelson, A~E. Jennings, and K~Kashima.
\newblock An earthquake history derived from stratigraphic and microfossil
  evidence of relative sea-level change at {C}oos {B}ay, southern coastal
  {O}regon.
\newblock \emph{Geological Society of America Bulletin}, 108\penalty0
  (2):\penalty0 141--154, 1996.

\bibitem[Parnell and Gehrels(2015)]{ParnellandGehrels}
A~C. Parnell and W~R. Gehrels.
\newblock Using chronological models in late {H}olocene sea-level
  reconstructions from salt marsh sediments.
\newblock In I.~Shennan, A.~J. Long, and B.~P. Horton, editors,
  \emph{{H}andbook of {S}ea-Level {R}esearch}, book section~32, pages 500--513.
  John Wiley \& Sons, 2015.

\bibitem[Parnell et~al.(2011)Parnell, Buck, and Doan]{Parnell2011}
A~C. Parnell, C~E. Buck, and T~K. Doan.
\newblock A review of statistical chronology models for high-resolution,
  proxy-based {H}olocene palaeoenvironmental reconstruction.
\newblock \emph{Quaternary Science Reviews}, 30\penalty0 (21-22):\penalty0
  2948--2960, 2011.

\bibitem[Parnell et~al.(2015)Parnell, Sweeney, Doan, Salter-Townshend, Allen,
  Huntley, and Haslett]{Bclim}
A~C. Parnell, J~Sweeney, T~K. Doan, M~Salter-Townshend, J~R.~M. Allen,
  B~Huntley, and J~Haslett.
\newblock {B}ayesian inference for palaeoclimate with time uncertainty and
  stochastic volatility.
\newblock \emph{Journal of the Royal Statistical Society: Series C (Applied
  Statistics)}, 64:\penalty0 115--138, 2015.

\bibitem[Plater et~al.(2015)Plater, Kirby, Boyle, Shaw, and Mills]{Plater2015}
A~J. Plater, J~R. Kirby, J~F. Boyle, T~Shaw, and H~Mills.
\newblock Loss on ignition and organic content.
\newblock In I.~Shennan, A.~J. Long, and B.~P. Horton, editors,
  \emph{{H}andbook of {S}ea-Level {R}esearch}, pages 312--330. John Wiley \&
  Sons, 2015.

\bibitem[Plummer(2003)]{JAGS}
M~Plummer.
\newblock {JAGS}: A program for analysis of {Bayesian} graphical models using
  {Gibbs} sampling.
\newblock In \emph{{P}roceedings of the 3rd {I}nternational {W}orkshop on
  {D}istributed {S}tatistical {C}omputing}, 2003.

\bibitem[Rasmussen(2006)]{Rasmussen06GP}
C~E. Rasmussen.
\newblock Gaussian {P}rocesses for {M}achine {L}earning.
\newblock MIT Press, 2006.

\bibitem[Roe et~al.(2002)Roe, Charman, and Gehrels]{Roe2002}
H~M. Roe, D~J. Charman, and R~W. Gehrels.
\newblock Fossil testate amoebae in coastal deposits in the {UK}: implications
  for studies of sea-level change.
\newblock \emph{Journal of Quaternary Science}, 17\penalty0 (5-6):\penalty0
  411--429, 2002.

\bibitem[Rymer(1978)]{Rymer1978}
L~Rymer.
\newblock The use of uniformitarianism and analogy in palaeoecology,
  particularly pollen analysis.
\newblock In D~Walker and J~C. Guppy, editors, \emph{Biology and Quaternary
  Environments}, pages 245--257. Australian Academy of Sciences, Canberra,
  1978.

\bibitem[Scott and Medioli(1978)]{Scott78}
D~B. Scott and F~S. Medioli.
\newblock Vertical zonations of marsh foraminifera as accurate indicators of
  former sea levels.
\newblock \emph{Nature}, 272\penalty0 (5653):\penalty0 528--531, 1978.

\bibitem[Shennan et~al.(1994)Shennan, Innes, Long, and Zong]{Shennan94}
I~Shennan, J~B. Innes, A~J. Long, and Y~Zong.
\newblock Late {D}evensian and {H}olocene relative sealevel changes at {L}och
  nan {E}ala, near {A}risaig, northwest {S}cotland.
\newblock \emph{Journal of Quaternary Science}, 9\penalty0 (3):\penalty0
  261--283, 1994.

\bibitem[Shennan et~al.(1996)Shennan, Rutherford, Innes, and
  Walker]{Shennan1996}
I~Shennan, M~M. Rutherford, J~B. Innes, and K~J. Walker.
\newblock Late glacial sea level and ocean margin environmental changes
  interpreted from biostratigraphic and lithostratigraphic studies of isolation
  basins in northwest {S}cotland.
\newblock \emph{Geological Society, London, Special Publications}, 111\penalty0
  (1):\penalty0 229--244, 1996.

\bibitem[Smith(1983)]{Smith83}
G~Smith.
\newblock \emph{{Quantitative {P}lant {E}cology}}, page 130.
\newblock University of California Press, 1983.

\bibitem[ter Braak and Juggins(1993)]{ter1993}
C~J.F. ter Braak and S~Juggins.
\newblock Weighted averaging partial least squares regression ({WA-PLS}): an
  improved method for reconstructing environmental variables from species
  assemblages.
\newblock \emph{Hydrobiologia}, 269-270\penalty0 (1):\penalty0 485--502, 1993.

\bibitem[Tingley et~al.(2012)Tingley, Craigmile, Haran, Li, Mannshardt, and
  Rajaratnam]{Tingley2012}
M~P. Tingley, P~F. Craigmile, M~Haran, B~Li, E~Mannshardt, and B~Rajaratnam.
\newblock Piecing together the past: statistical insights into paleoclimatic
  reconstructions.
\newblock \emph{Quaternary Science Reviews}, 35:\penalty0 1 -- 22, 2012.

\bibitem[Toivonen et~al.(2000)Toivonen, Mannila, Korhola, and
  Olander]{Toivonen2000}
H~T.~T. Toivonen, H~Mannila, A~Korhola, and H~Olander.
\newblock Applying {B}ayesian statistics to organism-based environmental
  reconstruction.
\newblock \emph{Ecological Applications}, 11\penalty0 (2):\penalty0 618--630,
  2000.

\bibitem[Tolwinski-Ward(2015)]{Ward2015}
S.~E. Tolwinski-Ward.
\newblock Uncertainty quantification for a climatology of the frequency and
  spatial distribution of {N}orth {A}tlantic tropical cyclone landfalls.
\newblock \emph{Journal of Advances in Modeling Earth Systems}, 7\penalty0
  (1):\penalty0 305--319, 2015.

\bibitem[Tolwinski-Ward et~al.(2013)Tolwinski-Ward, Anchukaitis, and
  Evans]{Tolwinski2013}
S~E. Tolwinski-Ward, K~J. Anchukaitis, and M~N. Evans.
\newblock Bayesian parameter estimation and interpretation for an intermediate
  model of tree-ring width.
\newblock \emph{Climate of the Past}, 9:\penalty0 1481--1493, 2013.

\bibitem[Tolwinski-Ward et~al.(2015)Tolwinski-Ward, Tingley, Evans, Hughes, and
  Nychka]{Tolwinski2015}
S.E. Tolwinski-Ward, M.P. Tingley, M.N. Evans, M.K. Hughes, and D.W. Nychka.
\newblock Probabilistic reconstructions of local temperature and soil moisture
  from tree-ring data with potentially time-varying climatic response.
\newblock \emph{Climate Dynamics}, 44\penalty0 (3-4):\penalty0 791--806, 2015.

\bibitem[Vasko et~al.(2000)Vasko, Toivonen, and Korhola]{Bummer}
K~Vasko, H~T.T. Toivonen, and A~Korhola.
\newblock A {B}ayesian multinomial {G}aussian response model for organism-based
  environmental reconstruction.
\newblock \emph{Journal of Paleolimnology}, 24\penalty0 (3):\penalty0 243--250,
  2000.

\bibitem[Wright et~al.(2011)Wright, Edwards, and van~de Plassche]{Wright2011}
A~J. Wright, R~J. Edwards, and O~van~de Plassche.
\newblock Reassessing transfer-function performance in sea-level reconstruction
  based on benthic salt-marsh foraminifera from the {A}tlantic coast of {NE}
  {N}orth {A}merica.
\newblock \emph{Marine Micropaleontology}, 81\penalty0 (1-€"2):\penalty0 43 --
  62, 2011.

\bibitem[Zong and Horton(1999)]{Zong1999}
Y~Zong and B~P. Horton.
\newblock Diatom-based tidal-level transfer functions as an aid in
  reconstructing {Q}uaternary history of sea-level movements in the {UK}.
\newblock \emph{Journal of Quaternary Science}, 14\penalty0 (2):\penalty0
  153--167, 1999.

\end{thebibliography}

\end{document}